\documentclass[a4paper, 12pt, openbib ]{article}
\usepackage{times}
\usepackage{pifont}
\usepackage{amsmath}
\usepackage{amssymb}
\usepackage{amsfonts}
\usepackage{graphicx}
\usepackage{floatflt}
\usepackage{geometry}
\usepackage{layout}
\usepackage{enumerate}
\usepackage{bm}
\usepackage{tikz}
\usetikzlibrary{patterns}
\def\ner{\boldsymbol}
\usepackage{amsthm}
\usepackage{thmtools}
\def\tfract#1/#2{{\textstyle{\raise0.8pt\hbox{$\scriptstyle#1$}\over%
\hbox{\lower0.8pt\hbox{$\scriptstyle#2$}}}}}
\def\mezzo{\tfract 1/2 }
\def\pidue{\tfract {\pi}/2}
\def\piqua{\tfract {\pi}/4}

\def\dueterzi{\tfract 2/3}

\def\radi2k{\tfract 1/{\sqrt {2k}} }
\def\der{\partial }
\def\cvd{\vbox{\hrule \hbox to 9 pt {\vrule height 9 pt \hfil \vrule} \hrule}}
\def\downnormalfill{$\,\,\vrule depth4pt width0.4pt
\leaders\vrule depth 0pt height0.4pt\hfill\vrule depth4pt width0.4pt\,\,$}
\def\WT#1{\mathop{\vbox{\ialign{##\crcr\noalign{\kern3pt}
      \downnormalfill\crcr\noalign{\kern0.8pt\nointerlineskip}
      $\hfil\displaystyle{#1}\hfil$\crcr}}}\limits}

\def\be{\begin{equation}}
\def\ee{\end{equation}}
\def\bes{\begin{equation*}}
\def\ees{\end{equation*}}
\def\bea{\begin{eqnarray}}
\def\eea{\end{eqnarray}}
\def\beas{\begin{eqnarray*}}
\def\eeas{\end{eqnarray*}}
\def\ba{\begin{array}{rcl}}
\def\ea{\end{array}}
\def\der{\partial}
\numberwithin{equation}{section}
\def\go{\leavevmode \raise.3ex\hbox{$\scriptscriptstyle \langle\!\langle\!  $}%
~\ignorespaces}
\def\gf{\relax \ifhmode \unskip~\else \leavevmode \fi \raise.3ex\hbox{$\! \scriptscriptstyle\rangle\!\rangle\, $}}
\title{
{\Large  Flat connections in three-manifolds \\ 
and classical Chern-Simons invariant}
{\vskip 0.5 truecm}}

\author{{\large  Enore~Guadagnini$^{\, a}$, Philippe Mathieu$^{\, b}$ and Frank~Thuillier$^{\, b}$} \\  {\normalsize {~}} \\  {\normalsize  $^{ a\, }$Dipartimento di Fisica {\it E. Fermi} dell'Universit\`a di Pisa, and INFN Sezione di Pisa,} \\ {\normalsize   Largo B. Pontecorvo  2, 56127 Pisa, Italy.}
 \\ {\normalsize $^{b\, }$LAPTh, Universit\'e de Savoie, CNRS, Chemin de Bellevue, BP 110,} \\ {\normalsize     F-74941 Annecy-le-Vieux Cedex, France.}
}
\date{}

\begin{document}

\maketitle

\vskip 0.7 truecm

\begin{abstract}

A general method for the construction of smooth  flat connections on 3-manifolds is introduced. The procedure  is strictly connected  with the deduction of the fundamental group of a manifold $M$ by means of a   Heegaard splitting presentation of  $M$. For any given   matrix  representation of the fundamental group of $M$, a corresponding flat connection $A$ on $M$ is  specified.   It is shown that the associated classical Chern-Simons invariant  assumes then a canonical form which is given by the sum of two contributions: the first term is determined by the intersections of the curves in the Heegaard diagram, and the second term  is the volume  of a region in the representation group which is determined by the representation of $\pi_1(M)$ and by the Heegaard gluing homeomorphism.  Examples of flat connections in topologically nontrivial manifolds are presented and the  computations of the associated  classical Chern-Simons invariants are illustrated.  

\end{abstract}

\vskip 1 truecm

\section {Introduction} Each $SU(N)$-connection, with $N \geq 2$, in a closed and oriented 3-manifold $M$ can be represented by a 1-form $A= A_\mu dx^\mu$ which takes  values in the  Lie algebra of  $SU(N)$.  The  Chern-Simons function $S[A]$,
\bea
 S[A] &=& \int_M {\cal L}_{CS} (A) =
  {1\over 8 \pi^2} \int_M {\rm Tr} \left ( A \wedge  d A + i \dueterzi A \wedge A \wedge A \right ) \nonumber \\ &=&
  {1\over 8 \pi^2} \int_M  d^3x \, \epsilon^{\mu \nu \lambda }  \, {\rm Tr} \left ( A_\mu (x) \der_\nu A_\lambda (x)+ i \dueterzi A_\mu (x) A_\nu (x)A_\lambda (x) \right )  \;  ,
\label{1.1}
\eea
can be   understood as the Morse function of an infinite dimensional Morse theory,   on which  the  instanton Floer homology \cite{F} and the gauge theory interpretation \cite{T} of the Casson invariant \cite{CL} are based. Under a local gauge transformation
\be
A_\mu (x) \longrightarrow A_\mu^\Omega (x) = \Omega^{-1} (x) A_\mu (x) \Omega (x) -i \Omega^{-1} (x)\der_\mu \Omega (x) \; ,
\label{1.2}
\ee
 where $\Omega $ is a map from $ M $ into $ SU(N)$, the   function $S[A]$ transforms as
 \be
 S[A^\Omega] = S[A] + I_\Omega \; ,
 \label{1.3}
 \ee
 where the integer $I_\Omega \in {\mathbb Z}$,
 \be
 I_\Omega = {1\over 24 \pi^2} \int_M {\rm Tr} \left ( \Omega^{-1} d \Omega \wedge \Omega^{-1} d \Omega \wedge \Omega^{-1} d \Omega \right ) \; ,
 \label{1.4}
 \ee
 can be used to label the homotopy class of $\Omega$.  The  stationary points of the function (\ref{1.1})  correspond to flat connections, {\it i.e.}  connections with vanishing curvature $F(A) = 2 dA + i [A, A] =0$. We shall now concentrate on flat connections exclusively.   Let $A$ be a flat  connection in $M$, and let  $\gamma \subset M$ be an oriented path  connecting the starting point $x_0$ to the final point $x_1$. The associated holonomy $\gamma \rightarrow h_\gamma [A] \in SU(N)$ is given by the path-ordered integral
\be
h_\gamma [A] = {\rm P } \, e^{i \int_\gamma A}  \; ,
\label{1.5}
\ee
which is computed along $\gamma$.  Under a   gauge transformation $A \rightarrow A^\Omega$, one finds
 \be
 h_\gamma [A^\Omega] = \Omega^{-1} (x_0) \,  h_\gamma [A] \, \Omega (x_1) \; .
 \label{1.6}
 \ee
Let us consider the set of holonomies  which are associated with the closed oriented paths  such that $x_0 = x_1 = x_b$, for a given base point $x_b$. Since the element $h_\gamma [A]\in SU(N)$  is invariant under homotopy transformations acting on $\gamma$,   this set of holonomies  specifies a matrix representation of the fundamental group $\pi_1(M)$ in the group $SU(N)$.  Because of equation (\ref{1.3}),  the classical Chern-Simons invariant $ cs [A]$,
 \be
 cs[A] = S[A] \quad \hbox{mod ~}   {\mathbb Z}   \; ,
 \label{1.7}
 \ee
is well defined  for the gauge orbits of flat $SU(N)$-connections on $M$, and it is well defined \cite{S} for the $SU(N)$ representations of $\pi_1 (M)$ modulo the action of  group conjugation. If the orientation of $M$ is modified, one gets $cs[A] \rightarrow - cs[A]$.  

 In the case of the structure group $SU(2)$,  methods for the computation  of $cs [A]$   have been  presented in References \cite{FI,KK,KKA, KKB,AU,AUX}, where a few non-unitary gauge groups have also been considered.  In all the examples that have been examined, $cs[A]$ turns out to be  a rational number.  In the case of three dimensional hyperbolic geometry, the associated $PSL(2 , {\mathbb C})$  classical invariant \cite{KKA,YO,DUP,JM} combines the real volume  and imaginary Chern-Simons parts in a complex geometric invariant. The Baseilhac-Benedetti invariant \cite{BB} with group $PSL(2 , {\mathbb C})$ represents some kind of  corresponding quantum invariant. 

 Precisely because flat connections represent  stationary points of the function (\ref{1.1}), flat connections and  the corresponding value of $cs[A]$ play an important role in the  quantum Chern-Simons gauge field theory \cite{W}.    For instance, the path-integral solution of the abelian Chern-Simons theory has recently been produced \cite{GT1,GT2}. In this case, flat connections dominate the functional integration and the value of the partition function is given by the sum over the gauge orbits of flat connections of  the exponential of the classical Chern-Simons invariant. The classical abelian Chern-Simons invariant is 
 strictly related  \cite{GT1,GT2} with the intersection quadratic form on the torsion group of $M$, which also enters the abelian Reshetikhin-Turaev  \cite{RT,MO} surgery invariant.   
 
 In general, the precise expression of the flat connections is  an essential ingredient for the  computation of the observables of the quantum  Chern-Simons theory by means of the path-integral  method. In this article we shall mainly be interested in  nonabelian flat connections.  
We will show that,  given  a representation $\rho $ of $\pi_1(M)$ and a Heegaard splitting presentation  \cite{R} of $M$  (with the related Heegaard diagram), by means of a general construction one can define a corresponding smooth flat connection $A$ on $M$. The method that we describe is  related   with the deduction \cite{FOX} of a presentation of the fundamental group of a manifold $M$ by means of a   Heegaard splitting of  $M$. Then the associated invariant $cs[A]$ assumes a canonical form, which can be written as  the sum of two contributions. The first term is determined by the intersections of the curves in the Heegaard diagram and can be interpreted as a sort of  ``coloured intersection form". Whereas the second term  is the Wess-Zumino volume  of a region in the structure  group $SU(N)$ which is determined by the representation of $\pi_1(M)$ and by the Heegaard gluing homeomorphism.  
 
The procedure that we present for the determination of the flat connections can find possible applications also in the description of the topological states of matter \cite{AS,BE}. 
A discussion on the importance of topological configurations and of the holonomy operators in  gauge theories can be found for instance in Ref.\cite{TH}. 

Our article is organised as follows. 
Section~2  contains a brief description of the main results of the present article. The general construction of flat connections in a generic 3-manifold $M$ by means of a Heegaard splitting  presentation of $M$ is discussed in Section~3. The canonical form of the corresponding classical Chern-Simons invariant is derived in Section~4, where a two dimensional formula   of the Wess-Zumino group volume is also produced.    In the remaining sections, our method is illustrated by a few examples. Flat connections in lens spaces are discussed in Section~5 and a non-abelian representation of the fundamental group of a particular 3-manifold is considered in Section~6; computations of the corresponding classical Chern-Simons invariants are presented. The case of the Poincar\'e sphere is  discussed in Section~7. One example of a general formula of the classic Chern-Simons invariant for a particular class of Seifert manifolds is given in Section~8. Finally, Section~9 contains the conclusions.  

 \section {Outlook}
 The main steps of our construction can be summarised as follows.  For any given $SU(N)$ representation $\rho$ of $\pi_1 (M)$,
 \be
\rho : \pi_1(M) \rightarrow SU(N) \; ,
\label{2.1}
\ee
 one can find a corresponding flat connection $A$ on $M$ whose   structure  is determined  by a Heegaard splitting presentation $M = H_L \cup_f H_R$ of $M$. In this presentation, the manifold $M$ is interpreted as the union of  two handlebodies $H_L$ and $H_R$ which are glued by means of the homeomorphism $f : \der H_L \rightarrow \der H_R$ of their boundaries, as sketched in Figure~1. 
 
 \vskip 0.6 truecm

\centerline {
\begin{tikzpicture} [scale=0.7] [>=latex]
%
\draw [very thick] (0,0) ellipse (1.55 and 1); 
\draw [very thick] (3,0) ellipse (1.55 and 1);
\fill[white] (1.1,-1) rectangle (1.9, +1);
\draw [very thick] plot  [smooth]  coordinates {(1.1,0.7)(1.3, 0.6)(1.5,0.56)(1.7,0.6)(1.9,0.7)}; 
\draw [very thick] plot  [smooth]  coordinates {(1.1,-0.7)(1.3, -0.6)(1.5,-0.56)(1.7,-0.6)(1.9,-0.7)}; 
\draw [very thick] (0,0.19)  ellipse (0.8 and 0.4);
\fill[white] (-1,0.1) rectangle (1, 0.7);
\draw [very thick] (0.66,-0.03) arc (0:180:0.66 and 0.3);
\draw [very thick] (3,0.19)  ellipse (0.8 and 0.4);
\fill[white] (2,0.1) rectangle (4, 0.7);
\draw [very thick] (3.66,-0.03) arc (0:180:0.66 and 0.3);
\draw [very thick] (10,0) ellipse (1.55 and 1); 
\draw [very thick] (13,0) ellipse (1.55 and 1);
\fill[white] (11.1,-1) rectangle (11.9, +1);
\draw [very thick] plot  [smooth]  coordinates {(11.1,0.7)(11.3, 0.6)(11.5,0.56)(11.7,0.6)(11.9,0.7)}; 
\draw [very thick] plot  [smooth]  coordinates {(11.1,-0.7)(11.3, -0.6)(11.5,-0.56)(11.7,-0.6)(11.9,-0.7)}; 
\draw [very thick] (10,0.19)  ellipse (0.8 and 0.4);
\fill[white] (9,0.1) rectangle (11, 0.7);
\draw [very thick] (10.66,-0.03) arc (0:180:0.66 and 0.3);
\draw [very thick] (13,0.19)  ellipse (0.8 and 0.4);
\fill[white] (12,0.1) rectangle (14, 0.7);
\draw [very thick] (13.66,-0.03) arc (0:180:0.66 and 0.3);
\draw [-> , very thick] (5.3,0.5) to [out=30 , in=150 ] (7.8,0.5);
\node at (6.55,1.5) {$  f $};
\node at (1.5,-1.8) {$  H_L $};
\node at (11.5,-1.8) {$  H_R $};
\end{tikzpicture}
}

\vskip 0.3 truecm
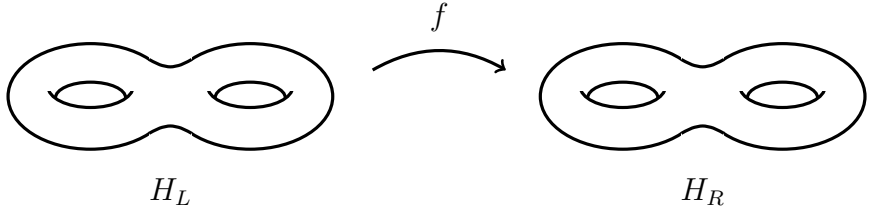
\centerline {{Figure 1.} {Attaching homemorphism $f : \der H_L \rightarrow \der H_R$.}}

\vskip 0.6 truecm
 
 Let  the fundamental group of $M$ be defined with respect to a base point $x_b$ which belongs to the boundaries of the two handlebodies. Then the representation $\rho $ of $\pi_1(M)$ canonically defines a representation of the fundamental group of each of the two handlebodies $H_L$ and $H_R$.  As shown in Figure~2, in each handlebody the generators of its fundamental group can be related with a set of corresponding disjoint meridinal discs. To each meridinal disc  is associated a matrix which is specified by the representation $\rho $; this matrix can be interpreted as a ``colour'' which is attached to each meridinal disc. With the help of these colored meridinal discs, one can construct a smooth flat connection $A^0_L$  in $H_L$ ---and similarly a smooth flat connection $A^0_R$ in $H_R$--- whose holonomies correspond to the elements of  the  representation $\rho $ in the handlebody $H_L$ (or $H_R$).  The precise definition of  $A^0_L$  and  $A^0_R$ is given in Section~3. 
 
 In general, $A^0_L$ and $A^0_R$ do not coincide with the restrictions in $H_L$ and $H_R$ of a single connection $A$ in $M$, because the images ---under $f$--- of the boundaries of the meridinal discs of $H_L$  are not the boundaries of meridinal discs of $H_R$. So, in order to define a connection $A$ which is globally defined in $M$, one needs to combine  $A^0_L$ with $A_R^0$ in a suitable way.  In facts, the exact matching of the gauge fields $A^0_L$ and $A^0_R$ in $M$  is specified by the  homeomorphism $f$ through the Heegaard diagram, which shows precisely how the boundaries of the meridinal discs of $H_L$ are pasted onto the surface $\der H_R$, in which the boundaries of the meridinal discs of $H_R$ are also placed. Let us denote by $f * A^0_L$ the image of $A^0_L$ under  $f$. The crucial point now is that,  on the surface $\der H_R$, the connections $A^0_R $ and $f * A^0_L$   are gauge related
 \be
 f* A^0_L = U_0^{-1} A^0_R U_0 -i  U_0^{-1} d U_0 \quad , \quad \hbox{ on~  }  \der H_R \; ,
 \label{2.2}
 \ee
because their holonomies define the same representation  of $\pi_1 ( \der H_R) $.  The value of the map $U_0$  from the surface $\der H_R$ on the group $SU(N)$  is uniquely determined by equation (\ref{2.2}) and by the condition $U_0(x_b) = 1$. In facts, we will demonstrate that
\be
U_0(x) = \Phi_R^{-1}(x) \Phi_{f* L}(x) \quad , \quad \hbox{ for ~~} x \in \der H_R \; ,
\label{2.3}
 \ee
 where $\Phi_R$ and $\Phi_{f* L}$ denote the developing maps associated respectively with $A^0_R$ and $f * A^0_L$ from the universal covering of $\der H_R $ into the group $SU(N)$. The definition of the developing map will be briefly recalled in Section~3.3.   Then the map $U_0$ can smoothly be extended  to the whole handlebody $H_R$; this extension will be denoted by $U$. The values of $U : H_R \rightarrow SU(N)$  inside $H_R$ are not constrained and can be chosen without restrictions apart from smoothness. As far as the computation of the classical Chern-Simons invariant is concerned,  the particular choice of  the extension $U$ of $U_0$  turns out to be irrelevant. To sum up, the connection $A$  ---which is well defined in $M$ and whose holonomies determine the representation $\rho $---   takes the form
  \be
 A  =  \left  \{ \begin{array}  {l@{ ~ } l}
  A^0_L & \quad \hbox{ in }  H_L \; ;    \\   ~ & ~ \\
 U^{-1}A_R^0 U -i U^{-1} d U &  \quad \hbox{ in }  H_R \; ;
\end{array} \right.
\label{2.4}
\ee
the correct matching of these two components is ensured by equation (\ref{2.2}). The expression (\ref{2.4}) of the connection implies

\newtheorem{prp}{Proposition}
\begin{prp}
The classical Chern-Simons invariant
{\rm (\ref{1.7})}, evaluated  for the $SU(N)$  flat connection {\rm (\ref{2.4})}, takes the form
\be
cs[A] = {\cal X}[A] + \Gamma [U] \quad \hbox{\rm mod ~}   {\mathbb Z} \; ,
\label{2.5}
\ee
where
\be
{\cal X} [A] =  {1 \over 8 \pi^2} \int_{\der H_R} {\rm Tr} \left [  U_0^{-1} A^0_R U_0 \wedge  f* A_L^0 \right ]\; ,
\label{2.6}
\ee
and
\be
\Gamma [U] =  {1 \over 24 \pi^2} \int_{H_R} {\rm Tr} \left [ U^{-1} d U \wedge  U^{-1} d U \wedge  U^{-1} d U\right ] \; .
\label{2.7}
\ee
\end{prp}

\noindent The function ${\cal X} [A]$ is defined on the surface $\der H_R$, and similarly the value of the Wess-Zumino volume $\Gamma [U]$~mod~${\mathbb Z}$  only depends \cite{WZ,N,WI} on  the values of $U$ in $\der H_R$ ({\it i.e.}, it only depends on $U_0$). A canonical dependence of $\Gamma$ on $U_0$ will be produced in Section~4.4. Therefore both terms in expression (\ref{2.5}) are  determined by the data on the two-dimensional surface $\der H_R$ of  the Heegaard splitting presentation $M = H_L \cup_f H_R$ exclusively. This is why the particular choice of the extension of $U_0$ inside $H_R$  is  irrelevant. The remaining part of this article contains the  proof of Proposition~1 and  a detailed  description of the construction of the flat connection $A$. Examples  will also be given, which elucidate the general procedure and illustrate the   computation of $cs[A] $. 

\section {Flat connections}
 Given a matrix representation $\rho $ of $\pi_1(M)$, we would like to determine a corresponding  flat connection $A$ on $M$ whose holonomies agree with $\rho $; then we shall compute $S[A]$.  
 
   In order to present a canonical construction which is not necessarily related with the properties of the representation space, we shall use a Heegaard splitting presentation $M = H_L \cup_f H_R$ of $M$.  The construction of $A$ is made of two steps.
First,  in each of the two handlebodies $H_L$ and $H_R$ we define a flat connection, $A^0_L$ and $A^0_R$ respectively, whose holonomies coincide with the elements of the matrix representation of  the fundamental group of the handlebody which is induced by $\rho $.
Second, the components $A^0_L$ and $A^0_R$ are combined according to the Heegaard diagram to define $A$ on $M$.

\subsection{Heegaard splitting}

Let us recall  \cite{S,R} that the fundamental group  of a three-dimensional  oriented handlebody $H$ of genus  $g$ is a free group with  $g$ generators  $\{ \gamma_1 , \gamma_2,..., \gamma_g \}$. 
 A disc $D$ in $H $ is called a meridinal disc if the boundary of $D$ belongs to the boundary of $H$, $\der D \subset \der H$,   and   $\der D$ is homotopically trivial in $H$.  Let $\{ D_1 , D_2 ,..., D_g\}$ be a set of disjoint meridinal discs in $H$   such that $H - \{ D_1 , D_2 ,..., D_g\}$ is homeomorphic with a 3-ball with $2g$ removed disjoint discs in its boundary. These meridinal discs $\{ D_1 , D_2 ,..., D_g\}$ can be put in a one-to-one correspondence with the $g$ handles of the handlebody $H$ or, equivalently,  with the generators of $\pi_1(H)$, and  can be oriented in such a way that the intersection of $\gamma_j$ with $D_k$ is $\delta_{jk}$. For instance, in the case of a handlebody of genus 2, a possible choice of the generators $\{ \gamma_1 , \gamma_2 \} $ and of the discs $\{D_1 , D_2 \}$ is illustrated in Figure~2, where the base point $x_b$ is also shown. 
 
  \vskip 0.75 truecm

\centerline {
\begin{tikzpicture} [scale=1.3] [>=latex]
%
\draw [very thick] (0,0) ellipse (1.55 and 1);
\draw [very thick] (3,0) ellipse (1.55 and 1);
\fill[white] (1.1,-1) rectangle (1.9, +1);
\draw [very thick] plot  [smooth]  coordinates {(1.09,0.71)(1.3, 0.6)(1.5,0.56)(1.7,0.6)(1.91,0.71)};
\draw [very thick] plot  [smooth]  coordinates {(1.09,-0.71)(1.3, -0.6)(1.5,-0.56)(1.7,-0.6)(1.91,-0.71)};
\draw [very thick] (0,0.19)  ellipse (0.8 and 0.4);
\fill[white] (-1,0.1) rectangle (1, 0.7);
\draw [very thick] (0.66,-0.03) arc (0:180:0.66 and 0.3);
\draw [very thick] (3,0.19)  ellipse (0.8 and 0.4);
\fill[white] (2,0.1) rectangle (4, 0.7);
\draw [very thick] (3.66,-0.03) arc (0:180:0.66 and 0.3);
%
\draw[very thick] (0,-1) arc (270:90:0.2 and 0.39);
\draw[very thick , densely dashed] (0,-1) arc (-90:90:0.2 and 0.39);
\draw[very thick] (3,-1) arc (270:90:0.2 and 0.39);
\draw[very thick , densely dashed] (3,-1) arc (-90:90:0.2 and 0.39);
\draw [thick] (0,0) ellipse (1.1 and 0.6);
\draw [thick] (3,0) ellipse (1.1 and 0.6);
\fill[white] (0.9,-0.4) rectangle (2.1, +0.4);
\draw [thick] plot  [smooth]  coordinates {(0.881,0.358)(1.25, 0.10)(1.5,0)(1.75,-0.10)(2.111,-0.352)};
\draw [thick] plot  [smooth]  coordinates {(0.881,-0.358)(1.25, -0.10)(1.5,0)(1.75,0.10)(2.111,0.352)};
\node at (0,-1.4) {$  D_1 $};
\node at (-1.3,0) {$\gamma_1$};
\node at (3,-1.4) {$  D_2$};
\node at (4.3,0) {$\gamma_2$};
\node at (1.5,0) {$\bullet$};
\node at (1.5,0.3) {$x_b$};
\end{tikzpicture}
}

\vskip 0.3 truecm
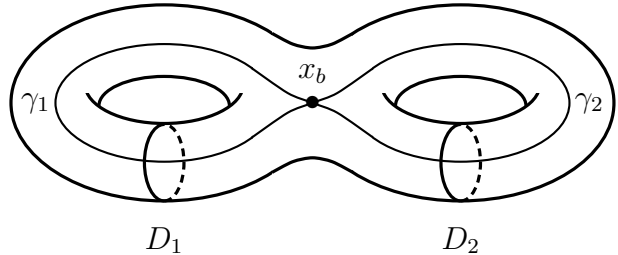
\centerline {{Figure~2.} {Generators $\{ \gamma_1 , \gamma_2 \}$ and meridinal discs $\{ D_1 , D_2 \}$  in a handlebody of genus 2.}}

\vskip 0.7 truecm

By means of a Heegaard  presentation $M = H_L \cup_f H_R$ of the 3-manifold $M$, which is specified by the homeomorphism 
\be
f \; : \; \der H_L \; \rightarrow \; \der H_R \; , 
\label{3.1}
\ee
one can find a presentation of  the  fundamental group $\pi_1(M)$. Suppose that the two handlebodies $H_L$ and $H_R$ have genus $g$. Let $\{ D_1 , D_2 ,..., D_g\}$ be a set of disjoint meridinal discs in $H_L$ which are associated with the $g$ handles of $H_L$. 
The homeomorphism $f : \der H_L \rightarrow \der H_R$  is specified ---up to ambient isotopy---  by  the images $ C^\prime_j = f (C_j)$  in $\der H_R$ of the boundaries $ C_j = \der D_j $, for $j =1,2,.., g$.    Thus each Heegaard splitting can be described by a diagram which shows  the set  of the characteristic curves $\{ C^\prime_j \} $ on the surface $\der H_R$.   
One example of Heegaard diagram is shown in Figure~3. 

 \vskip 0.65 truecm

\centerline {
\begin{tikzpicture} [scale=1.6] [>=latex]
%
\draw [very thick] (0,0) ellipse (1.55 and 1);
\draw [very thick] (3,0) ellipse (1.55 and 1);
\fill[white] (1.1,-1) rectangle (1.9, +1);
\draw [very thick] plot  [smooth]  coordinates {(1.09,0.71)(1.3, 0.6)(1.5,0.56)(1.7,0.6)(1.91,0.71)};
\draw [very thick] plot  [smooth]  coordinates {(1.09,-0.71)(1.3, -0.6)(1.5,-0.56)(1.7,-0.6)(1.91,-0.71)};
\draw [very thick] (0,0.19)  ellipse (0.8 and 0.4);
\fill[white] (-1,0.1) rectangle (1, 0.7);
\draw [very thick] (0.66,-0.03) arc (0:180:0.66 and 0.3);
\draw [very thick] (3,0.19)  ellipse (0.8 and 0.4);
\fill[white] (2,0.1) rectangle (4, 0.7);
\draw [very thick] (3.66,-0.03) arc (0:180:0.66 and 0.3);
\draw [thick ] (0,0) ellipse (1.1 and 0.6);
\node at (0,-.415) {$C^\prime_1$};
\draw [thick ] (0,0) ellipse (1.4 and 0.8);
\draw [thick ] (3,0) ellipse (1.4 and 0.8);
\fill[white] (1.15,-0.45) rectangle (1.85, +0.45);
\draw [thick ]  (1.15,0.455) .. controls (1.3,0.33) and (1.55,0.25) .. (1.85,0.455); 
\draw [thick ] (3,0) ellipse (1.1 and 0.6);
\fill[white] (1.82,-0.41) rectangle (2.1, -0.63);
\fill[white] (2.2,-0.39) rectangle (2.8, -0.66);
\fill[white] (2.78,-0.69) rectangle (3.2, -0.5);
\fill[white] (2.08,-0.78) rectangle (2.8, -0.6);
\fill[white] (2.58,-0.71) rectangle (3.5, -0.9);
\fill[white] (3.48,-0.83) rectangle (3.7, -0.6);
\draw [thick ]  (1.15,-0.455) .. controls (1.3,-0.33) and (1.55,-0.25) .. (1.85,-0.455); 
\draw [thick ]  (1.83,-0.44) .. controls (2.3,-0.75) and (3,-0.602) .. (3.2,-0.59); 
\draw [thick ]  (2.18,-0.4) .. controls (2.48,-0.55) and (2.7,-0.2) .. (2.9,-0.21); 
\draw [thick ]  (3.2,-1) .. controls (3.4,-0.93) and (3.37,-0.8) .. (3.71,-0.69); 
\draw [thick , dashed]  (2.9,-0.21) .. controls (3.1,-0.28) and (3,-0.93) .. (3.2,-1); 
\node at (1.68,0) {$C^\prime_2$};
\end{tikzpicture}
}

\vskip 0.4 truecm
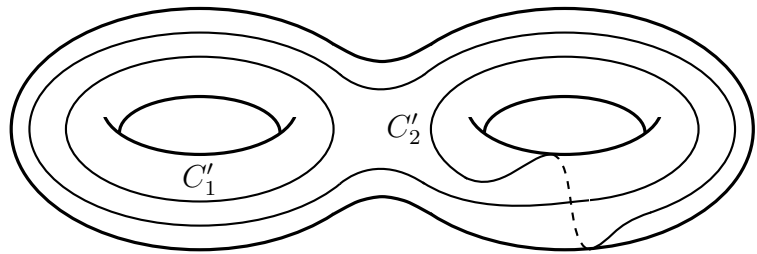
\centerline {{Figure~3.} {Example of a genus 2 Heegaard diagram.}}

\vskip 0.5 truecm

\noindent Let $\{ \gamma_1, \gamma_2,..., \gamma_g \}$ be a complete set of generators for $\pi_1(H_R)$ which are associated to a complete set of  meridinal discs of $H_R$.    The fundamental group of $M$ is specified  by adding to the generators 
$\{ \gamma_1, \gamma_2,..., \gamma_g \}$ the constraints which implement the homotopy triviality condition of the curves $\{ C^\prime _j \} $. Indeed, since each curve $C_j$ is homotopically trivial in $M$,  the fundamental group of $M$ admits \cite{R,FOX} the presentation 
 \be
 \pi_1(M) = \langle \, \gamma_1, \gamma_2,..., \gamma_g \, | \, [C_1^\prime ] =1  ,...,   [C_g^\prime ] =1 \, \rangle \; , 
 \label{3.2}
 \ee
 where $[C^\prime_j ]$ denotes the $\pi_1 (H_R)$  homotopy class of $C^\prime_j$  expressed in terms of the generators $\{ \gamma_1, \gamma_2,..., \gamma_g \}$.  The classes $[C^\prime_j ]$ are determined by the intersections of the boundaries of the meridinal discs of $H_L$ and $H_R$, which can be inferred from the Heegaard diagram. 
 
\subsection{Flat connection in a handlebody} 

Let us consider the handlebody $H_L$ of the Heegard splitting $M = H_L \cup_f H_R$ of genus $g$ and a corresponding set  $\{ D_1 , D_2 ,..., D_g\}$  of disjoint meridinal discs in $H_L$.  For each $j = 1,2,..., g$, consider   a collared neighbourhood $N_j $  of  $D_j$  in $H_L$.  As shown in Figure~4, $N_j$ is homeomorphic with a cylinder $D_j \times [0,\epsilon ] $ parametrized as $(z\in {\mathbb C} , |z| \leq 1 ) \times ( 0\leq t \leq \epsilon )$.

\vskip 0.1 truecm

\centerline {
\begin{tikzpicture} [scale=0.7] [>=latex]
%
\draw [ very thick ] (-5,0) ellipse (1 and 2);
\node at (-4.70,0) {$D_j$};
\draw [  -> , very thick  ] (-3,0) to (-2,0);
\draw[ very thick ] (0,2) arc (90:270:1 and 2);
\draw[ very thick , densely dashed] (0,-2) arc (-90:90:1 and 2);
\draw [ very thick ] (1,0) ellipse (1 and 2);
\draw [ very thick  ] (0,2) to (1,2);
\draw [ very thick ] (0,-2) to (1,-2);
\node at (1.53,0) {$D_j$};
\node at (0.5,2.4) {$t$};
\end{tikzpicture}
}

\vskip 0.4 truecm
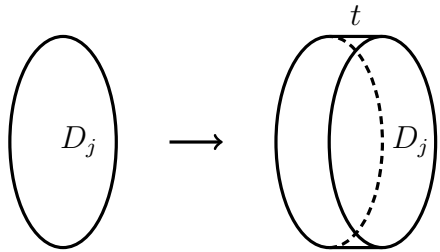
\centerline {{Figure~4.} {Disc $D_j$ and the neighbourhood $N_j $  of  $D_j$.}}

\vskip 0.5 truecm

\noindent The strip $(|z|=1)\times ( 0\leq t \leq \epsilon )$ belongs to the surface $\der H_L$. 
 The flat $SU(N)$-connection on $H_L$  we are interested in will be denoted by $A_L^0$; $A_L^0$  is vanishing in $H_L - \{ N_1 , N_2,..., N_g \}$ and,  inside each region $N_j$, $A_L^0$ is determined by $\rho (\gamma_j)$. More precisely, suppose that
\be
\rho(\gamma_j) = e^{i b_j}\; ,
\label{3.3}
\ee
where the hermitian traceless matrix $b_j$ belongs to the Lie algebra of $SU(N)$. Let $\theta (t)$ be a ${\cal C}^\infty $ real function, with $ \theta^\prime (t) = d \theta (t)/ dt  > 0 $, satisfying $\theta (0)= 0 $ and $\theta (\epsilon ) = 1$. Then the value of $A_L^0$ in the region $N_j$ is given by
\be
A_L^0 \Big |_{N_j} = b_j \theta^\prime (t) dt \; .
\label{3.4}
\ee
The orientation of the parameterization (or the sign in equation (\ref{3.4})) is fixed so that the holonomy of the connection (\ref{3.4}) coincides with expression (\ref{3.3}). 
As a consequence of equation (\ref{3.4})  one has $d A^0_L = 0$ and also, since  $N_j  \cap N_k = \emptyset $ for $j \not= k$, one finds $A^0_L \wedge A^0_L =0 $.

  By construction, the smooth 1-form $A_L^0$ represents a flat connection on $H_L$ whose holonomies coincide with  the  matrices that represent the  elements of the fundamental group of $H_L$. The restriction of $A_L^0$ on the boundary $\der H_L$ has support on   $g$ ribbons and its values are determined by equation (\ref{3.4}); the $j$-th ribbon represents a collared neighbourhood of the curve $C_j = \der D_j$  in $\der H_L$.  The same construction can be applied to define a flat connection $A^0_R$ on   $H_R$.

\subsection{Flat connection in a 3-manifold}

Let us now  construct a flat connection $A$  in $M = H_L \cup_f H_R$ which is  associated with the representation $\rho$ of $\pi_1(M)$. As far as  the value   of $A$ on $H_L$ is concerned, one can put
\be
A \Big |_{H_L} = A^0_L \; .
\label{3.5}
\ee
The image $f* A^0_L$ of $A^0_L$ under the homeomorphism $f : \der H_L \rightarrow \der H_R$ does not coincide in general with  $A_R^0$ in $\der H_R$. But since $f* A^0_L$ and $A_R^0$ are associated with the same matrix representation of $\pi_1(\der H_R)$,  the values of $f* A^0_L$ and $A_R^0$ on $\der H_R$ are related by a gauge transformation, $ f* A^0_L = U_0^{-1} A^0_R U_0 -i  U_0^{-1} d U_0$,  as shown in equation (\ref{2.2}), in which $U_0$ must assume the unit value at the base point $x_b$.  Then the map $U_0$ can smoothly be extended in $H_R$, let $U$ denote this extension. The value of $A$ on $H_R$ is taken to be
\be
A \Big |_{H_R} = U^{-1}A_R^0 U -i U^{-1} d U \; .
\label{3.6}
\ee
The value  of $U_0$  on the surface $\der H_R$ represents  a fundamental ingredient of our construction, so we now describe how  it  can be determined.   To this end, we need to introduce the concept of developing map.

 Let us recall that any flat $SU(N)$-connection $A$ defined in a space $X$ can be  locally trivialized because, inside  a simply connected neighbourhood of any given  point of $X$,   $A$ can be written as $A = - i \Phi^{-1} d \Phi $. The value  of  $\Phi $ coincides with the holonomy of $A$. When the representation of $\pi_1 (X)$ determined by $A$ is not trivial, $\Phi$ cannot be extended to the whole space $X$.  A global trivialisation of $A$ can be found in the universal covering $\widehat X$ of $X$; in this case, the map $\Phi : \widehat X \rightarrow SU(N) $ represents the developing map. For any element $\gamma $ of $\pi_1 (X)$ acting on $ \widehat X$ by covering transformations, the developing map satisfies
 \be
 \Phi(\gamma \cdot x) = h_\gamma [A] \cdot  \Phi( x) \; ,
 \label{3.7}
  \ee
 in agreement with equations (\ref{1.6}).  Now, on the surface $\der H_R$ we have the two flat connections $f* A^0_L$  and $ A^0_R$ which are related by a gauge transformation, equation (\ref{2.2}).    Thus, for each oriented path $\gamma \subset \der H_R$ connecting the starting point $x_0$ with the final point $x$, the corresponding holonomies are related according to equation (\ref{1.6}) which takes the form
 \be
U_0^{-1}(x_0)\, h_\gamma [ A^0_R] \, U_0(x) =  h_\gamma [f* A^0_L] \; .
\label{3.8}
\ee
 From this equation one obtains $
 U_0(x) = h^{-1}_\gamma [A^0_R] \, U_0 (x_0)\, h_\gamma [f* A^0_L] $.
When the starting point $x_0$  coincides with the base point  $x_b$ of the  fundamental group,   one has $U (x_b) =1$, and then
 \be
U_0(x) = h^{-1}_\gamma [ A^0_R] \,\, h_\gamma [f* A^0_L] \quad , \quad \hbox{ for ~~} x \in \der H_R  \; .
 \label{3.9}
\ee
This equation is equivalent to the relation (\ref{2.3}). Indeed, because of the transformation property (\ref{3.7}), the combination $\Phi^{-1}_R \Phi_{f* L}$ is invariant under covering translations acting on the universal covering of $\der H_R$ (and then $\Phi^{-1}_R \Phi_{f* L}$ is really a map from $\der H_R$ into $SU(N)$),  and locally coincides with the product $h^{-1}_\gamma [ A^0_R] \,h_\gamma [f*  A^0_L]$ appearing in equation (\ref{3.9}).

\section {The invariant}

\subsection{Proof of Proposition~1}

The Chern-Simons function $S[A]$ of the connection (\ref{2.4}) ---whose components in $H_L$ and $H_R$ are shown in equations (\ref{3.5}) and (\ref{3.6})--- is given by
\be
S[A] = \int_M {\cal L}_{CS} (A) = \int_{H_L} {\cal L}_{CS} (A) + \int_{H_R} {\cal L}_{CS} (A)\; .
\label{4.1}
\ee
Since $dA_L^0 =0$ and $A_L^0 \wedge A^0_L =0 $, one has
\be
 \int_{H_L} {\cal L}_{CS} (A) = \int_{H_L} {\cal L}_{CS} (A^0_L) = 0 \; .
 \label{4.2}
 \ee
Moreover, a direct computation shows that
\bea
\int_{H_R} {\cal L}_{CS} (A) &=& \int_{H_R} {\cal L}_{CS} (A^0_R) - {i\over 8 \pi^2} \int_{ H_R} d \, {\rm Tr} \left [  A^0_R \wedge d U U^{-1}\right ]\nonumber \\
&& + {1 \over 24 \pi^2} \int_{H_R} {\rm Tr} \left [ U^{-1} d U \wedge  U^{-1} d U \wedge  U^{-1} d U\right ] \; .
\label{4.3}
\eea
As before, the first term on the r.h.s of equation (\ref{4.3}) is vanishing
\be
\int_{H_R} {\cal L}_{CS} (A^0_R) =0 \; .
\label{4.4}
\ee
By using equation (\ref{2.2}), the second term can be written as the surface integral
\be
{\cal X} [A] =  {1 \over 8 \pi^2} \int_{\der H_R} {\rm Tr} \left [  U_0^{-1} A^0_R U_0 \wedge  f* A_L^0 \right ]\; .
\label{4.5}
\ee
By combining equations (\ref{4.1})-(\ref{4.5}) one finally gets 
\bea
S[A] &=& {1 \over 8 \pi^2} \int_{\der H_R} {\rm Tr} \left [  U_0^{-1} A^0_R U_0\wedge  f* A_L^0 \right ]\nonumber \\ 
&& + {1 \over 24 \pi^2} \int_{H_R} {\rm Tr} \left [ U^{-1} d U \wedge  U^{-1} d U \wedge  U^{-1} d U\right ] \; , 
\label{4.6}
\eea
which implies equation (\ref{2.5}). This concludes the proof of Proposition~1.

The  term ${\cal X} [A]$ can be understood as a sort of colored intersection form, because  its value is determined by the trace of the representation matrices ---belonging to the Lie algebra of the group--- which are associated  with the boundaries of the meridinal discs of the two handlebodies which intersect each other in the Heegaard diagram.  Indeed, on the surface $\der H_R$,  $A^0_R$ is different from zero inside collar neighbourhoods of the boundaries of the meridinal discs of $H_R$, whereas  $f* A_L^0$ is different from zero inside collar neighbourhoods of the images ---under $f$--- of the boundaries of the meridinal discs of $H_L$. Thus, in the computation of ${\cal X} [A] $, only the intersection regions of the curves of the Heegaard diagram give nonvanishing  contributions.  But since the intersections of the boundaries of the meridinal discs of $H_L$ and $H_R$ determine the relations entering the presentation (\ref{3.2})  of $\pi_1 (M)$,  an important  part of the input, which is involved in the computation of ${\cal X}[A]$,  is given by  the fundamental group presentation (\ref{3.2}).  
It turns out that the computation of ${\cal X} [A] $ can also be accomplished  by means of intersection theory techniques by coloring the de Rham-Federer  currents \cite{CUR1,CUR2} of the disks $\{ D_j \}$. 

When the representation $\rho $ is abelian,  $\Gamma [U]$ vanishes and the classical Chern-Simons invariant is completely specified by ${\cal X}[A]$ which assumes  the simplified  form    
\be
cs[A] \, \Big |_{abelian}  = {\cal X} [A] \, \Big |_{abelian} =  {1 \over 8 \pi^2} \int_{\der H_R} {\rm Tr} \left [  A^0_R  \wedge  f* A_L^0 \right ]\quad \hbox{\rm mod ~}   {\mathbb Z} \; .
\label{4.7}
\ee

\subsection{Group volume} 

The  term $\Gamma [U]$ can be interpreted as the 3-volume  of the  region of the structure group which is bounded by the image of the surface $\der H_R$ under the map $\Phi^{-1}_R \,\Phi_{f* L}  : \der H_R \rightarrow SU(N)$. 
In this case also, the combination $\Phi^{-1}_R \,\Phi_{f* L} $ of the two developing maps, which are associated with  $f*A^0_L$ and $A^0_R$, is characterized   by the homeomorphism $f : \der H_L \rightarrow \der H_R$ which topologically identifies $M$. 

In general, the direct computation of $\Gamma [U]$ is not trivial, and the following properties of 
$\Gamma [U]$ turns out to be useful. When $U(x)$ can be written as 
\be
U(x) = W(x) \, Z(x) \; , 
\label{4.8}
\ee
where $W(x) \in SU(N)$ and $Z(x)\in SU(N)$, 
one obtains  
\be
\Gamma [U = WZ] = \Gamma [W]
 + \Gamma [Z] + {1 \over 8 \pi^2} \int_{\der H_R} {\rm Tr} \left [  d Z Z^{-1} \wedge W^{-1} d W    \right ] \; \; . 
 \label{4.9}
 \ee
 By means of equation (\ref{4.9}) one can easily derive  the relation 
 \bea
 \Gamma [U = V H V^{-1}] &=& \Gamma [H] - {1 \over 8 \pi^2} \int_{\der H_R} {\rm Tr} \left [ V^{-1} d V \wedge \left ( H^{-1} d H + d H H^{-1}\right ) \right ] \nonumber \\
 &&+ {1 \over 8 \pi^2} \int_{\der H_R} {\rm Tr} \left [ V^{-1} d V  \, H \, \wedge V^{-1} d V H^{-1} \right ] \; . 
 \label{4.10}
 \eea
With a clever choice of the matrices $V(x)$ and $H(x)$, equation (\ref{4.10}) assumes a simplified form.  Indeed any generic map $U(x) \in SU(N)$ can locally be written in the form $U(x) = V(x) H(x) V^{-1}(x)$ where 
\be 
H(x) = \exp ( i C(x)) \; , 
\label{4.11}
\ee
 and $C(x)$ belongs to the $(N-1)$-dimensional  abelian Cartan  subalgebra of the Lie algebra of $SU(N)$. In this case, one has $\Gamma[H] =0 $ and 
 \be 
 H^{-1} (x) d H (x)= dH (x)\, H^{-1}(x) = i\,  d C(x) \; . 
 \label{4.12}
 \ee
Therefore relation (\ref{4.10}) becomes 
  \be
 \Gamma [  V H V^{-1}] =  {1\over 8 \pi^2} \int_{\der H_R} \Bigl \{ 2 i \, {\rm Tr} \left [ dC \, \wedge V^{-1} d V \right ] +  {\rm Tr} \left [e^{-i C} \, V^{-1} d V  \, e^{i C} \, \wedge V^{-1} d V  \right ] \Bigr \} \; ,  
 \label{4.13}
 \ee
 where it is understood that one possibly needs to decompose the integral into a sum of integrals computed  in different regions of $\der H_R$ where $V(x)$ and $H(x)$ are well defined \cite{GAW}.   Expression (\ref{4.13}) explicitly shows that the value of $ \Gamma [U]$ (modulo integers) is completely specified by the value of $U$ on the surface $\der H_R$. 
 
In the case of the structure group $SU(2) \sim S^3$, the computation of $\Gamma[U]$ can be reduced to the computation of the volume of a given polyhedron in a space of constant curvature. Discussions on this last problem can be found, for instance,   in the articles \cite{SC,V,CH,DMP,MURA,YSK,AGM,AB}.

 \subsection {Canonical extension}

The reduction of the  Wess-Zumino volume $\Gamma [U] $ into a  surface integral on $\der H_R$ can be done in several inequivalent ways, which also depend on the choice of the  extension   of $U_0 $ from the surface $\der H_R$ in $H_R$. Let us now describe the result which is obtained by means of a canonical extension of $U_0$. We shall concentrate on the structure group $SU(2)$, the generalisation to a  generic  group $SU(N)$  is  quite simple.   

 Suppose that the value of $U_0$ on the surface $\der H_R$ can be written as 
\bea
U_0 (x,y) &=& e^{ i \ner n (x,y) \ner \sigma} = e^{ i  \sum_{a=1}^3 n^a (x,y) \, \sigma^a}
 \nonumber \\ 
 &=&   \cos n(x,y) + i \, \widehat {\ner  n} (x,y) \ner \sigma \, \sin n (x,y)  \; , 
\label{4.14}
\eea
where $(x,y)$ designate coordinates of $\der H_R$,   $ n = \left [ \sum_{b=1}^3n^b n^b \right ]^{1/2} $, the components of the unit vector $\widehat {\ner n}$ are  given by $\widehat {\ner n}^a = n^a / n $,  and $\{ \sigma^a \} $ (with $a=1,2,3$) denote the Pauli sigma matrices. 
The canonical extension of $U_0$ is defined by 
\be
U (\tau, x,y) = e^{ i \, \tau \,  \ner n (x,y) \ner \sigma} \; ,
\label{4.15}
\ee
where the homotopy parameter $\tau $ takes values in the range $ 0 \leq \tau \leq 1$. When $\tau = 1 $ one recovers the expression (\ref{4.14}), whereas in the $\tau \rightarrow 0 $ limit one finds $U =1$.  A direct computation gives
\be
 {\rm Tr} \left ( U^{-1} \der _\tau U \, \left [ U^{-1}\der_x U, U^{-1}\der_y U\right ] \right ) =  {2 i\over n^2} \sin^2 (\tau n )  \, {\rm Tr} \left (    \Sigma \, \left [ \der_y \Sigma \, , \der_x \Sigma \right ]  \right )\; , 
 \label{4.16}
\ee
in which $\Sigma(x,y) = \sum_{a=1}^3n^a (x,y) \sigma^a$. 
Therefore, by using the identity 
\be
\int_0^1 d \tau  \, \sin^2 (\tau n ) =  \mezzo \left [ 1 - {\sin (2 n)\over 2 n}   \right ]\; , 
\label{4.17}
\ee
one gets 
\be 
\Gamma[U]  = {-i \over 8 \pi^2} \int_{\der H_R}
{1 \over n^2} \left [ 1 -  {\sin (2n) \over  2n}  \right ] \,  {\rm Tr} \left ( \Sigma \, d \Sigma \wedge d \Sigma \, \right ) \; .  
\label{4.18}
\ee
This equation will be used in Section~6,  Section~7 and Section~8.
 
\subsection{Rationality}
\label{RA}
As it has already been mentioned, in all the considered examples the value of the $SU(N)$ classical Chern-Simons invariant is given by a rational number. Let us now present a proof of this property for  a particular  class of 3-manifolds.    Suppose that the universal covering $\widetilde M$ of the three-manifold $M$ is homeomorphic with $S^3$,  so that $M$ can be identified with the orbit space  \cite{MA} which is obtained     by means of covering translations (acting on  $S^3 $) which correspond to the elements of the  fundamental group $\pi_1 (M)$.    Given a flat connection $A$ on $M$, let us denote by $\widetilde A$ the flat connection on $\widetilde M  \sim S^3$ which is the upstairs preimage of  $A$. By construction, one has 
\be
S[A] \Bigr |_{M} = {1 \over |\pi_1(M)|} \, S[\widetilde A] \, \Bigr |_{S^3} \; , 
\label{4.19}
\ee
where $|\pi_1(M)|$ denotes the order of $\pi_1 (M)$. On the other hand, since $S^3$ is simply connected, one can find a map $\Omega : S^3 \rightarrow SU(N)$  such that 
\be
\widetilde A  = -i \Omega^{-1} \, d \Omega 
\; , 
\label{4.20}
\ee
and then 
\be
S[\widetilde A] \, \Bigr |_{S^3} = {1\over 24 \pi^2} \int_{S^3} {\rm Tr} \left ( \Omega^{-1} d \Omega \wedge \Omega^{-1} d \Omega \wedge \Omega^{-1} d \Omega \right ) = n \; , 
\label{4.21}
\ee
where $n$ is an integer. Equations (\ref{4.19}) and (\ref{4.21}) imply 
\be
cs[ A ] \, \Bigr |_{M} = {n \over |\pi_1(M)|} 
\quad \hbox{\rm mod ~}   {\mathbb Z} \; ,  
\label{4.22}
\ee
which shows that, for this type of manifolds, the value of $cs [A]$ is indeed a rational number.  

Let us now present a few examples of computations of  $cs[A]$; in the first instance,  the  representation of the fundamental group of the 3-manifold  is abelian, whereas  nonabelian representations are considered in the remaining examples.    

 \section{First example}
In order to illustrate how to compute ${\cal X} [A]$, let us consider the lens spaces $L(p,q)$, where  the coprime integers $p$ and $q$ verify $p>1$ and $1 \leq q < p$. The manifolds $L(p,q)$  admit \cite{S,R} a genus~1 Heegaard splitting presentation,   $L(p,q) = H_L \cup_f H_R$ where $H_L$ and $H_R$ are solid tori. The fundamental group of   $L(p,q)$ is  the abelian group $\pi_1(L(p,q)) = {\mathbb Z}_p$. 

 \vskip 0.5 truecm

\centerline {
\begin{tikzpicture} [scale= 0.7] 
%
\draw [very thick  ] (1,1) circle (1.7); 
\draw [very thick  ] (8,1) circle (1.7); 
\draw [  very thick
] (1.9,2.45) to [out=20 , in=160 ] (7.11,2.45);
\draw [  very thick
] (2.5,1.8) to [out=15 , in=165 ] (6.5,1.8);
\draw [  very thick
] (2.7,1) to [out=0 , in=180 ] (6.3,1);
\draw [  very thick
] (2.5,0.2) to [out=-15 , in=-165 ] (6.5,0.2);
\draw [  very thick
] (1.9,-0.45) to [out=-20 , in=-160 ] (7.11,-0.45);
\node at (1,1) {$  + F $};
\node at (8,1) {$  - F $};
\node at (1.6,2.1) {\footnotesize $  3 $};
\node at (2.1,1.7) {\footnotesize $  4 $};
\node at (2.3,1) {\footnotesize $  5 $};
\node at (2.1,0.3) {\footnotesize $  1 $};
\node at (1.6,-0.1) {\footnotesize $  2 $};
\node at (7.4,2.1) {\footnotesize $  1 $};
\node at (6.9,1.7) {\footnotesize $  2 $};
\node at (6.7,1) {\footnotesize $  3 $};
\node at (6.9,0.3) {\footnotesize $  4 $};
\node at (7.4,-0.1) {\footnotesize $  5 $};
\node at (3.2,1.5) {$\bullet$};
\node at (3.8,1.5) {$x_b$};
\node at (6,-1.3){$C^\prime$}; 
\draw [-> , very thick ] (-0.2,-0.2) --  +(132:1mm);  
\draw [-> , very thick ] (4.5,2.97) -- ++(180:1mm); 
\draw [-> , very thick ] (4.5,2.104) -- ++(180:1mm); 
\draw [-> , very thick ] (4.5,1) -- ++(180:1mm); 
\draw [-> , very thick ] (4.5,-0.103) -- ++(180:1mm);  
\draw [-> , very thick ] (4.5,-0.97) -- ++(180:1mm); 
\draw [-> , very thick ] (9.7,1) -- ++(90:1mm); 
\end{tikzpicture}
}

\vskip 0.3 truecm
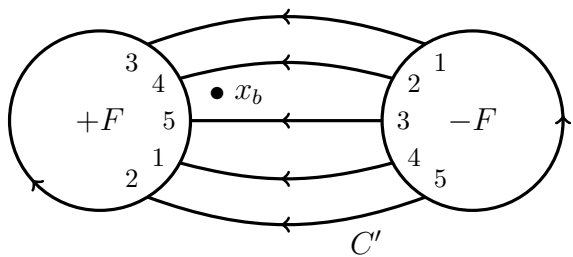
\centerline {{Figure~5.} {Heegaard diagram for the lens space $L(5,2)$, with base point $x_b$ displayed.}}

\vskip 0.3 truecm

\subsection{The representation} Let us concentrate, for example, on $L(5,2)$ whose  Heegaard diagram is shown in Figure~5, where the image $C^\prime$ of a meridian  $C$ of the solid torus $H_L$ is displayed on the surface  $\der H_R$.  The torus $\der H_R$ is represented by the surface of a 2-sphere with two removed discs $+F$ and $-F$. The  boundaries of $+F$ and $-F$ must be identified (the points with the same label coincide). A possible choice of the base point $x_b$ of the fundamental group is also depicted. 

In the solid torus $H_L$, let the meridian  $C$ be the boundary of the meridinal disc $D_L\subset H_L$, which is oriented so that the intersection of $D_L$ with the  generator $\gamma_L \subset H_L$ of $\pi_1(H_L)$ is +1.  Suppose that the representation $\rho : \pi_1 (L(5,2)) = {\mathbb Z}_5 \rightarrow SU(4)$ is specified by  
\be
\rho (\gamma_L ) = \exp \left [  i {2 \pi  \over 5} Y\right ] \; , 
\label{5.1}
\ee
where  $Y$ is given by 
\be
Y =  \begin{pmatrix} 1 & 0 & 0 & 0\\ 0 & 1 & 0 & 0\\ 0 & 0 & 1 & 0 \\ 0 & 0 & 0 & -3 \end{pmatrix} \; . 
\label{5.2}
\ee
  Let $N_L\subset H_L$ be a collared neighbourhood of $D_L$ parametrised by $(z\in {\mathbb C} , |z| \leq 1 ) \times ( 0\leq t \leq \epsilon )$. The flat connection $A^0_L$ on $H_L$ is vanishing in $H_L - N_L$, whereas the value of $A^0_L $ in $N_L$ is given by 
\be
A^0_L \, \Big |_{N_L} =   {2 \pi  \over 5} Y \theta^\prime (t) dt \; . 
\label{5.3}
\ee
The restriction  of $A^0_L$ on the boundary $\der H_L$ is nonvanishing  inside a strip which is a collared neighbourhood of $C$. Therefore the image $f* A^0_L$ of $ A^0_L$ on $\der H_R$ is different from zero in a collared neighbourhood of $C^\prime$. 

Let us now consider $H_R$. The meridinal disc $D_R \subset H_R $ can be chosen in such a way that the boundary of $D_R$ coincides with the boundaries of $+F$ (and $-F$) of Figure~5. The image  on $\der H_R$ of the corresponding generator  $\gamma_R$ of $\pi_1(H_R)$     is  associated to $+F$, and it can be represented by an arrow intersecting the boundary of the disc $+F$  and oriented in the outward direction.  As in the previous case, we introduce a collared neighbourhood $N_R \subset H_R$ of $D_R$ 
parametrised by $(z^\prime \in {\mathbb C} , |z^\prime | \leq 1 ) \times ( 0\leq u \leq \epsilon )$.
The flat connection $A^0_R$ is vanishing in $H_R - N_R$ and, inside $N_R$, one has  
\be
A^0_R \, \Big |_{N_R} =   \widetilde Y \,  \theta^\prime (u ) d u \; ,  
\label{5.4}
\ee
where  $\widetilde Y $  represents an element of the Lie algebra of $SU(N)$.  The restriction  of $A^0_R$ on the boundary $\der H_R$ is nonvanishing inside a collared neighbourhood of $\der (+ F)$.  The value taken by $A^0_R$ must be consistent with the given representation  $\rho : \pi_1 (L(5,2))  \rightarrow SU(4)$ which is specified by equation (\ref{5.1}). In order to determine $A^0_R$, one can  consider a closed path $\gamma \subset \der H_R$ with base point  $x_b$. One needs to impose that  the holonomy  of $A^0_R$ along $\gamma $  must coincide with the holonomy of $f* A^0_L$  along $\gamma $. One then finds  $\widetilde Y = (4\pi /5) Y$, and consequently  
\be
A^0_R \, \Big |_{N_R} =   {4 \pi  \over 5} Y \,  \theta^\prime (u ) d u \; .   
\label{5.5}
\ee
    
As shown in the Heegaard diagram of Figure~5,  the collar neighbourhood of $C^\prime $ and the collar neighbourhood of $\der (+F) $ ---where the connections $f*A^0_L$ and $A^0_R$ are nonvanishing---  intersect in five  (rectangular) regions of $\der H_R$.  Only inside  these  rectangular regions is the 2-form $A^0_R \wedge f*A^0_L$ different from zero. 
As far as the computation of the Chern-Simons  invariant is concerned, these five regions are equivalent and give the same contribution to ${\cal X}[A]$. The values of the connections inside one of the five rectangular intersection regions are shown in Figure~6.
 
\vskip 0.2 truecm

\centerline {
\begin{tikzpicture} [scale= 1.0] 
%
\draw [  very thick  , dashed ] (1,1) -- (6,1);
\draw [  very thick ] (1,5) -- (6,5);
\draw [ very thick ] (1,6) -- (1,0);
\draw [ very thick , dashed ] (5,0) -- (5,6);
\draw [ -> , thick ] (5.4, 4.9) -- (5.4, 1.1);
\node at (5.7, 4.7) {\small 0};
\node at (5.7, 1.25) {$\small {\epsilon }$};
\node at (5.7, 3) {$t$};
\draw [ -> , thick ] (1.1, 0.7) -- (4.9, 0.7);
\node at (1.3, 0.4) {\small 0};
\node at (4.7, 0.4) {$\small {\epsilon }$};
\node at (3, 0.4) {$u$};
\node at (3,5.4) {$ C^\prime $};
\node at (0.4,3) {$+F$};
\draw [-> , very thick ] (4,5) -- ++(180:1mm); 
\draw [-> , very thick ] (1,4) -- ++(-90:1mm); 
\node at (3,3.7) {$f*A^0_L=  {2 \pi  \over 5} Y \theta^\prime (t) dt$ };
\node at (3,2.3){$A^0_R =   {4 \pi  \over 5} Y \,  \theta^\prime (u ) d u $};
\end{tikzpicture}
}

\vskip 0.5 truecm

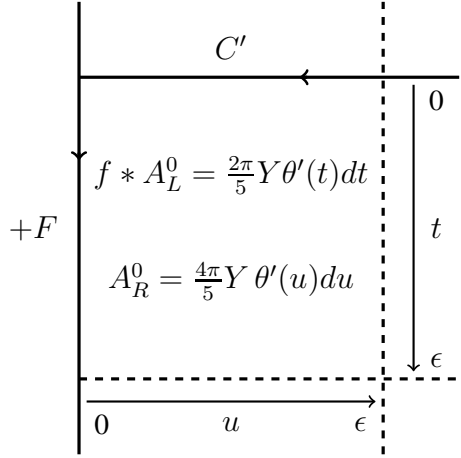
\centerline {{Figure~6.} {Values of the connections inside one intersection region.}}

\vskip 0.3 truecm

\noindent In the intersection region shown in Figure~6, one then finds 
\be
\int_{\footnotesize \hbox{one region}} {\rm Tr} \left [   A^0_R  \wedge  f* A_L^0 \right ]  = - {8 \pi^2 \over 25}\int_0^\epsilon dt \, \theta^\prime (t) \int_0^\epsilon du \, \theta^\prime (u) \, {\rm Tr} \left [ Y^2 \right ] = - {96 \pi^2 \over 25} \; . 
\label{5.6}
\ee
Therefore the value of the classical Chern-Simons  invariant  which, in this abelian case,  takes the form 
\be
cs[A] = {1 \over 8 \pi^2} \int_{\der H_R} {\rm Tr} \left [  A^0_R \wedge  f* A_L^0 \right ] \quad \hbox{\rm mod ~}   {\mathbb Z}\; ,
\label{5.7}
\ee
 is given by  
\be
cs[A] =  {5 \times \left ( - 96 \pi^2 / 25 \right )\over 8 \pi^2}  \quad \hbox{\rm mod ~}   {\mathbb Z} \; \; =  {3\over 5} \quad \hbox{\rm mod ~}   {\mathbb Z} \; .   
\label{5.8}
\ee

\subsection{Lens spaces in general}

For a generic lens space $L(p,q)$, the corresponding Heegaard diagram has the same structure of the diagram shown in Figure~5.  The curve $C^\prime $ on $\der H_R$ and the boundary of the disc $(+F) $ give rise to  $p$ intersection regions.  
Since the group $\pi_1(L(p,q))$ is abelian, the analogues of equations (\ref{5.3}) and (\ref{5.5}) take the form 
\be
A^0_L \, \Big |_{N_L} =   {2 \pi  \over p} M \theta^\prime (t) dt \; , 
\label{5.9}
\ee
and
\be
A^0_R \, \Big |_{N_R} =   {2 \pi  q \over p} M \,  \theta^\prime (u ) d u \; ,    
\label{5.10}
\ee
where the matrix $M$ belongs to the Lie algebra of $SU(N)$ and satisfies 
\be
e^{i 2 \pi M} = 1 \; . 
\label{5.11}
\ee
 Therefore the expression of the classical Chern-Simons invariant (\ref{5.7}) is given by  
\be
cs[A] = - {1 \over 8 \pi^2} \biggl \{  {(2 \pi )^2 \, q \over p^2}   \, {\rm Tr} \left ( M^2 \right )\times p   \biggr \} = - 
 {q \over p}\Bigl [ \, \mezzo  \, {\rm Tr} \left ( M^2 \right ) \Bigr ] \quad \hbox{\rm mod ~}   {\mathbb Z}\; . 
\label{5.12}
\ee
Expression (\ref{5.12}) is in agreement with the results \cite{GT1,GT2} obtained in the case of  the abelian Chern-Simons theory, where it has been shown that the value of the Chern-Simons action is specified by the quadratic intersection  form on the torsion component of the homology group of the manifold. 

\section{Second example}

Let us consider the 3-manifold $\Sigma_3$ which is homeomorphic with the cyclic 3-fold  branched covering of $S^3$ which is branched over the trefoil \cite{R}.  $\Sigma_3$ admits a Heegaard splitting presentation of genus 2 and the corresponding Heegaard diagram is shown in Figure~7. The surface $\der H_R$  is represented by the surface of a 2-sphere with four removed discs: the boundaries of $+F$ and $-F$ (and similarly the boundaries of $+G$ and $-G$) must be identified.  In Figure~7, the two characteristic curves $C^\prime_1$ and $C^\prime_2$ are represented  by the continuous and the dashed curve respectively, and the base point $x_b$ is also shown. 

 \vskip 0.5 truecm

\centerline {
\begin{tikzpicture} [scale= 0.8] 
%
\draw [very thick  ] (5,7) circle (1); 
\draw [very thick  ] (5,1) circle (1); 
\draw [very thick  ] (1,4) circle (1); 
\draw [very thick  ] (9,4) circle (1); 
\node at (1,4) {$  + F $};
\node at (9,4) {$  - F $};
\node at (5,1) {$  + G $};
\node at (5,7) {$  - G $};
\draw [  very thick
] (1.2,5) to [out=60 , in=190 ] (4,6.8);
\draw [  very thick , dashed 
] (2,4.2) to [out=10 , in=240 ] (4.8,6);
\draw [  very thick , dashed 
] (1.2,3) to [out=-60 , in=170 ] (4,1.2);
\draw [  very thick 
] (2,3.8) to [out=-10 , in=120 ] (4.8,2);
\draw [  very thick  
] (5.2,6) to [out=-60 , in=170 ] (8,4.2);
\draw [  very thick , dashed 
] (6,6.8) to [out=-10 , in=120 ] (8.8,5);
\draw [  very thick , dashed 
] (5.2,2) to [out=60 , in=190 ] (8,3.8);
\draw [  very thick
] (6,1.2) to [out=10 , in=240 ] (8.8,3);
%
%
%
\node at (1.1,4.7) {\footnotesize $  3 $};
\node at (1.7,4.2) {\footnotesize $  4 $};
\node at (1.7,3.8) {\footnotesize $  1 $};
\node at (1.1,3.3) {\footnotesize $  2 $};
\node at (4.25,6.9) {\footnotesize $  2 $};
\node at (4.72,6.3) {\footnotesize $  3 $};
\node at (5.25,6.3) {\footnotesize $  4 $};
\node at (5.7,6.9) {\footnotesize $  1 $};
\node at (4.25,1.2) {\footnotesize $  1 $};
\node at (4.69,1.71) {\footnotesize $  2 $};
\node at (5.25,1.71) {\footnotesize $  3 $};
\node at (5.7,1.2) {\footnotesize $  4 $};
\node at (8.9,3.3) {\footnotesize $  3 $};
\node at (8.3,3.8) {\footnotesize $  2 $};
\node at (8.3,4.2) {\footnotesize $  1 $};
\node at (8.9,4.7) {\footnotesize $  4 $};
\node at (2.6,2.5) {$\bullet$};
\node at (3.1,2.5) {$x_b$};
\node at (8.2,1.6) {$C_1^\prime$};
\node at (8.4,6.3) {$C_2^\prime$};
\draw [-> , very thick ] (2.5,6.315) --  +(214:1mm);  
\draw [-> , very thick ] (3.7,4.8) -- ++(208:1mm); 
\draw [-> , very thick ] (3.7,3.2) -- ++(330:1mm); 
\draw [-> , very thick ] (2.5,1.68) -- ++(145:1mm); 
\draw [-> , very thick ] (7.6,6.26) -- ++(147:1mm);  
\draw [-> , very thick ] (6.6,4.63) -- ++(335:1mm); 
\draw [-> , very thick ] (6.5,3.32) -- ++(216:1mm); 
\draw [-> , very thick ] (7.4,1.636) -- ++(207:1mm); 
\draw [-> , very thick ] (0.4,4.8) -- ++(38:1mm); 
\draw [-> , very thick ] (4.282,0.3) -- ++(135:1mm); 
\end{tikzpicture}
}

\vskip 0.4 truecm
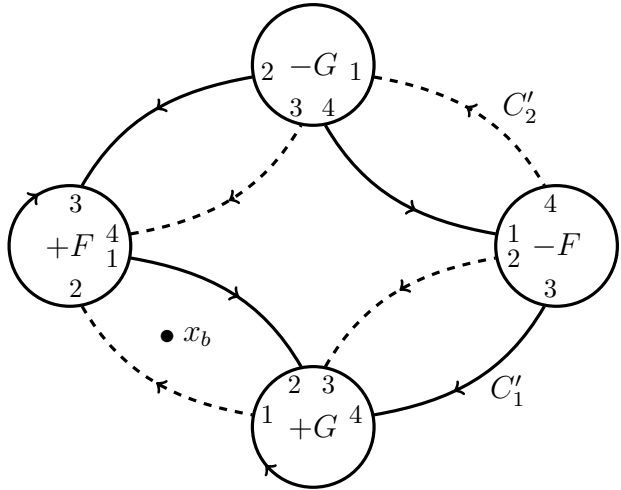
\centerline {{Figure~7.} {Heegaard diagram for $\Sigma_3$, with base point $x_b$ displayed.}}

\vskip 0.3 truecm

The two meridinal discs $D_{1R}$ and $D_{2R} $ of $H_R$ are chosen so that their boundaries coincide with the boundaries of the discs $+F$ and $+G$ respectively. The corresponding generators $\gamma_1$ and $\gamma_2$ of $\pi_1(H_R)$ can be  represented by two arrows which are based on the boundaries of $+F$ and $+G$ and oriented in the outward direction. By taking into account the constraints coming from the requirement of homotopy triviality of the curves $C^\prime_1$ and $C^\prime_2$, one finds a presentation of the fundamental group of $\Sigma_3$,  
\be
\pi(\Sigma_3 ) = \langle \gamma_1 \, \gamma_2 \, | \, \gamma_1^2 = \gamma_2^2 = (\gamma_1 \gamma_2 )^2 \, \rangle \; . 
\label{6.1}
\ee
The group $\pi(\Sigma_3 )$ is usually called \cite{R} the quaternionic group; it has eight elements which can be denoted by $\{ \pm 1 , \pm i , \pm j , \pm k \}$, in which $ij = k$, $ki=j$ and $jk=i$.  

Let the representation $\rho : \pi_1 (\Sigma_3) \rightarrow SU(2)$ be given by 
\bea
\gamma_1 &\rightarrow& g_1 = \exp \left [ i (\pi /2)\sigma^1 \, \right ] = i  \begin{pmatrix} 0 & 1 \\  1 & 0 \end{pmatrix} = i \sigma^1 \quad , \nonumber \\ 
\gamma_2 &\rightarrow& g_2 = \exp \left [ i (\pi /2)\sigma^2 \, \right ] = i  \begin{pmatrix} 0 & -i \\  i & 0 \end{pmatrix} = i \sigma^2 \quad . 
\label{6.2}
\eea
The corresponding flat connection $A^0_R$ on $H_R$ vanishes in $H_R - \{ N_{1R}  , N_{2R}\} $, where $N_{1R}$ and $N_{2R}$ are collared neighbourhoods of the two meridinal discs $\{ D_{1R} , D_{2R} \}$ of $H_R$, and  
 \be
A^0_R = \left  \{ \begin{array}  {l@{ ~ } l}   A^0_R  \, \Big |_{N_{1R}} =   { \pi  \over 2} \sigma^1\, 
 \theta^\prime (u) du & {\; \; }  ;   \\  ~ & ~ \\  
A^0_R  \, \Big |_{N_{2R}} =   { \pi  \over 2} \sigma^2 \, \theta^\prime (v) dv & {\; \; } .   
\end{array} \right. 
\label{6.3}
\ee
With the choice of the base point $x_b$ shown in Figure~7, the flat connection $A^0_L $ on $H_L$ turns out to be 
\be
A^0_L = \left  \{ \begin{array}  {l@{ ~ } l}   A^0_L  \, \Big |_{N_{1L}} =   { \pi  \over 2} \sigma^1\, 
 \theta^\prime (t) dt & {\; \; }  ;   \\  ~ & ~ \\  
A^0_L  \, \Big |_{N_{2L}} =   { \pi  \over 2} \sigma^2 \, \theta^\prime (s) ds & {\; \; } ;   
\end{array} \right. 
\label{6.4}
\ee
where $N_{1L}$ and $N_{2L}$ are collared neighbourhoods of the two meridinal discs $\{ D_{1L} , D_{2L} \}$ of $H_L$, and $A^0_L $ vanishes on $H_L - \{ N_{1L} , N_{2L}\} $. 
Note that, on the surface $\der H_R$,  $f*A^0_L$  is nonvanishing inside the two  ribbons which constitute collared neighbourhoods of the curve $C_1^\prime $ and $C^\prime_2$, whereas  $A^0_R$ is nonvanishing inside the  collared neighbourhoods of $\der D_{1R}$ and $\der D_{2R}$. In the region of the surface $\der H_R$ where both $f*A^0_L$ and $A^0_R$ are vanishing, the values taken by the map $U_0$ entering equation (\ref{2.2}) are shown in Figure~8.

 \vskip 0.4 truecm

\centerline {
\begin{tikzpicture} [scale= 0.8] 
%
\draw [very thick  ] (5,7) circle (1); 
\draw [very thick  ] (5,1) circle (1); 
\draw [very thick  ] (1,4) circle (1); 
\draw [very thick  ] (9,4) circle (1); 
\node at (1,4) {$  + F $};
\node at (9,4) {$  - F $};
\node at (5,1) {$  + G $};
\node at (5,7) {$  - G $};
\draw [  very thick
] (1.2,5) to [out=60 , in=190 ] (4,6.8);
\draw [  very thick , dashed 
] (2,4.2) to [out=10 , in=240 ] (4.8,6);
\draw [  very thick , dashed 
] (1.2,3) to [out=-60 , in=170 ] (4,1.2);
\draw [  very thick 
] (2,3.8) to [out=-10 , in=120 ] (4.8,2);
\draw [  very thick  
] (5.2,6) to [out=-60 , in=170 ] (8,4.2);
\draw [  very thick , dashed 
] (6,6.8) to [out=-10 , in=120 ] (8.8,5);
\draw [  very thick , dashed 
] (5.2,2) to [out=60 , in=190 ] (8,3.8);
\draw [  very thick
] (6,1.2) to [out=10 , in=240 ] (8.8,3);
%
%
%
\node at (1.1,4.7) {\footnotesize $  3 $};
\node at (1.7,4.2) {\footnotesize $  4 $};
\node at (1.7,3.8) {\footnotesize $  1 $};
\node at (1.1,3.3) {\footnotesize $  2 $};
\node at (4.25,6.9) {\footnotesize $  2 $};
\node at (4.72,6.3) {\footnotesize $  3 $};
\node at (5.25,6.3) {\footnotesize $  4 $};
\node at (5.7,6.9) {\footnotesize $  1 $};
\node at (4.25,1.2) {\footnotesize $  1 $};
\node at (4.69,1.71) {\footnotesize $  2 $};
\node at (5.25,1.71) {\footnotesize $  3 $};
\node at (5.7,1.2) {\footnotesize $  4 $};
\node at (8.9,3.3) {\footnotesize $  3 $};
\node at (8.3,3.8) {\footnotesize $  2 $};
\node at (8.3,4.2) {\footnotesize $  1 $};
\node at (8.9,4.7) {\footnotesize $  4 $};
\node at (3,2.5) {$1$};
\node at (4.98,4) {$g_1$};
\node at (7,2.55) {$g_1g_2$};
\node at (6.95,5.55) {$-1$};
\node at (3,5.55) {$g_2g_1$};
\node at (1.5,1.2) {$g_2$};
\draw [-> , very thick ] (2.5,6.315) --  +(214:1mm);  
\draw [-> , very thick ] (3.7,4.8) -- ++(208:1mm); 
\draw [-> , very thick ] (3.7,3.2) -- ++(330:1mm); 
\draw [-> , very thick ] (2.5,1.68) -- ++(145:1mm); 
\draw [-> , very thick ] (7.6,6.26) -- ++(147:1mm);  
\draw [-> , very thick ] (6.6,4.63) -- ++(335:1mm); 
\draw [-> , very thick ] (6.5,3.32) -- ++(216:1mm); 
\draw [-> , very thick ] (7.4,1.636) -- ++(207:1mm); 
\draw [-> , very thick ] (0.4,4.8) -- ++(38:1mm); 
\draw [-> , very thick ] (4.282,0.3) -- ++(135:1mm); 
\end{tikzpicture}
}

\vskip 0.4 truecm
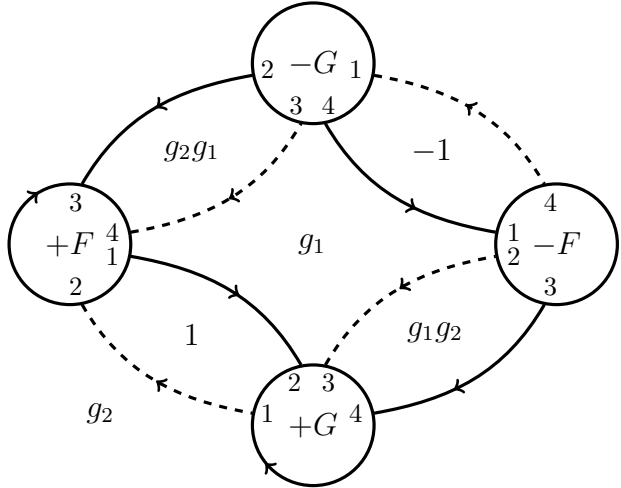
\centerline {{Figure~8.} {Values of the map $U_0$ in the region where $f*A^0_L$ and $A^0_R$ are vanishing.}}

\vskip 0.4 truecm

 We now need to specify  the values of $U_0 = \Phi_R^{-1}\Phi_{f*L}$  in the eight intersections regions of $\der H_R$ where both  $f*A^0_L$ and $A^0_R$ are not vanishing.   The value of $U_0$ is defined in equation  (\ref{3.9}).  In each region, we shall introduce the variables $X$ and $Y$ according to a correspondence of the type  
\bea
dX &=&  \theta^\prime (t) dt \quad , \quad 0 \leq X \leq 1 \nonumber \\ 
dY &=&  \theta^\prime (u) du \quad , \quad 0 \leq Y \leq 1 \; . 
\label{6.5}
\eea
The intersection regions are denoted as $\{ F1, F2, F3, F4, G1 , G2 , G3 , G4 \} $ with the convention that, for instance, the region $F3$ (or $G3$) is a rectangle in which one of the vertices is the point denoted by the number $3$ of the boundary of the disk $+F$ (or $+G$).  
The values of $U_0$ in these eight regions are in order; in each of the corresponding pictures, the values of $U_0$ at the vertices of the rectangle are also reported.

\centerline {
\begin{tikzpicture} [scale= 0.7] 
%
\draw [ -> ,  very thick  ] (9,0) -- (9,4);
\draw  (11,0) -- (11,4);
\draw [ -> ] (8,1) -- (12,1);
\draw [ very thick ] (8,3) -- (12,3); 
\node at (9.5, 4) {$Y$}; 
\node at (12.1, 1.4) {$X$}; 
\node at (10, 2) {$F1$};
\node at (8.5, 0.5) {$1$};
\node at (11.5, 0.5) {$g_1$};
\node at (11.5, 3.5) {$-1$};
\node at (8.5, 3.5) {$g_1$};
\node at (3,2){$\left [ F1\right ] : $ \qquad $U_0= e^{i \pidue ( X + Y )\sigma^1}$};
\end{tikzpicture}
}

\vskip 0.2 truecm

\centerline {
\begin{tikzpicture} [scale= 0.7] 
%
\draw [ ->   ] (9,0) -- (9,4);
\draw [ very thick ] (11,0) -- (11,4);
\draw [ -> , very thick ] (8,1) -- (12,1);
\draw  (8,3) -- (12,3); 
\node at (9.5, 4) {$Y$}; 
\node at (12.1, 1.4) {$X$}; 
\node at (10, 2) {$F2$};
\node at (8.5, 0.5) {$1$};
\node at (11.5, 0.5) {$g_1$};
\node at (11.7, 3.5) {$g_1 g_2$};
\node at (8.5, 3.5) {$g_2$};
\node at (3,2){$\left [ F2\right ] : $ \qquad $U_0= e^{i \pidue X \sigma^1} e^{i \pidue Y \sigma^2}$};
\end{tikzpicture}
}

\vskip 0.2 truecm

\centerline {
\begin{tikzpicture} [scale= 0.7] 
%
\draw [ -> ,  very thick  ] (9,0) -- (9,4);
\draw  (11,0) -- (11,4);
\draw [ -> , very thick ] (8,1) -- (12,1);
\draw  (8,3) -- (12,3); 
\node at (9.5, 4) {$Y$}; 
\node at (12.1, 1.4) {$X$}; 
\node at (10, 2) {$F3$};
\node at (8.3, 0.5) {$g_1 g_2$};
\node at (11.5, 0.5) {$g_2$};
\node at (11.7, 3.5) {$g_2 g_1$};
\node at (8.5, 3.5) {$g_2$};
\node at (3,2){$\left [ F3\right ] : $ \qquad $U_0= e^{i \pidue ( 1 - X - Y )\sigma^1} e^{i \pidue  \sigma^2}$};
\end{tikzpicture}
}

\vskip 0.2 truecm

\centerline {
\begin{tikzpicture} [scale= 0.7] 
%
\draw [ -> ,  very thick  ] (9,0) -- (9,4);
\draw  (11,0) -- (11,4);
\draw [ -> ] (8,1) -- (12,1);
\draw [ very thick ] (8,3) -- (12,3); 
\node at (9.5, 4) {$Y$}; 
\node at (12.1, 1.4) {$X$}; 
\node at (10, 2) {$F4$};
\node at (8.4, 0.5) {$-1$};
\node at (11.5, 0.5) {$g_1$};
\node at (11.7, 3.5) {$g_2 g_1$};
\node at (8.5, 3.5) {$g_2$};
\node at (3,2){$\left [ F4\right ] : $ \qquad $U_0= e^{-i \pidue  X \sigma^1} e^{i \pidue ( 2- Y) \sigma^2}  $};
\end{tikzpicture}
}

\vskip 0.2 truecm

\centerline {
\begin{tikzpicture} [scale= 0.7] 
%
\draw [ -> ,  very thick  ] (9,0) -- (9,4);
\draw  (11,0) -- (11,4);
\draw [ -> ] (8,1) -- (12,1);
\draw [ very thick ] (8,3) -- (12,3); 
\node at (9.5, 4) {$Y$}; 
\node at (12.1, 1.4) {$X$}; 
\node at (10, 2) {$G1$};
\node at (8.5, 0.5) {$1$};
\node at (11.5, 0.5) {$g_2$};
\node at (11.5, 3.5) {$-1$};
\node at (8.5, 3.5) {$g_2$};
\node at (3,2){$\left [ G1\right ] : $ \qquad $U_0= e^{i \pidue ( X + Y )\sigma^2}$};
\end{tikzpicture}
}

\vskip 0.2 truecm

\centerline {
\begin{tikzpicture} [scale= 0.7] 
%
\draw [ -> ] (9,0) -- (9,4);
\draw [ very thick ] (11,0) -- (11,4);
\draw [ -> , very thick ] (8,1) -- (12,1);
\draw (8,3) -- (12,3); 
\node at (9.5, 4) {$Y$}; 
\node at (12.1, 1.4) {$X$}; 
\node at (10, 2) {$G2$};
\node at (8.5, 0.5) {$1$};
\node at (11.5, 0.5) {$g_2$};
\node at (11.7, 3.5) {$g_2 g_1$};
\node at (8.5, 3.5) {$g_1$};
\node at (3,2){$\left [ G2\right ] : $ \qquad $U_0= e^{i \pidue X \sigma^2} e^{i \pidue Y \sigma^1}$};
\end{tikzpicture}
}

\vskip 0.2 truecm

\centerline {
\begin{tikzpicture} [scale= 0.7] 
%
\draw [ -> ,  very thick  ] (9,0) -- (9,4);
\draw (11,0) -- (11,4);
\draw [ -> , very thick ] (8,1) -- (12,1);
\draw  (8,3) -- (12,3); 
\node at (9.5, 4) {$Y$}; 
\node at (12.1, 1.4) {$X$}; 
\node at (10, 2) {$G3$};
\node at (8.3, 0.5) {$g_2 g_1$};
\node at (11.5, 0.5) {$g_1$};
\node at (11.7, 3.5) {$g_1 g_2$};
\node at (8.5, 3.5) {$g_1$};
\node at (3,2){$\left [ G3\right ] : $ \qquad $U_0= e^{i \pidue ( 1 - X - Y )\sigma^2} e^{i \pidue  \sigma^1}$};
\end{tikzpicture}
}

\vskip 0.2 truecm

\centerline {
\begin{tikzpicture} [scale= 0.7] 
%
\draw [ -> ,  very thick  ] (9,0) -- (9,4);
\draw  (11,0) -- (11,4);
\draw [ ->  ] (8,1) -- (12,1);
\draw [ very thick ] (8,3) -- (12,3); 
\node at (9.5, 4) {$Y$}; 
\node at (12.1, 1.4) {$X$}; 
\node at (10, 2) {$G4$};
\node at (8.4, 0.5) {$-1$};
\node at (11.5, 0.5) {$g_2$};
\node at (11.7, 3.5) {$g_1 g_2$};
\node at (8.5, 3.5) {$g_1$};
\node at (3,2){$\left [ G4\right ] : $ \qquad $U_0= e^{i \pidue  (2-X )\sigma^2} e^{-i \pidue  Y \sigma^1}  $};
\end{tikzpicture}
}

\vskip 0.2 truecm

\noindent  By using the value of $U_0$ in the eight intersections regions $\{ F1, F2, F3, F4, G1 , G2 , G3 , G4 \} $,  the contribution ${\cal X}[A]$, defined in equation (\ref{4.5}),  of the Chern-Simons invariant can easily be determined. One finds 
\bea 
{\cal X}[A] &=& {1\over 8 \pi^2} \, {\rm Tr} \, \Bigl \{ - \piqua \sigma^1 \sigma^1 + \piqua \sigma^1 \sigma^2 + \piqua \sigma^1 \sigma^1 + \piqua \sigma^1 \sigma^2 \nonumber \\ 
&& \qquad - \piqua \sigma^2 \sigma^2 + \piqua \sigma^2 \sigma^1 + \piqua \sigma^2 \sigma^2 + \piqua \sigma^2 \sigma^1  \Bigr \} = 0 \; . 
\label{6.6}
\eea
Let us now consider the computation of the contribution $\Gamma [A]$ of equation (\ref{2.7}).  Under the map $U_0 = \Phi_R^{-1} \Phi_{f*L} : \der H_R \rightarrow SU(2)$,  the images of the rectangles $\{ F1 , F3 , G1 , G3 \}$   are degenerate (they have codimension two). Whereas the images    of the remaining four rectangles $\{ F2 , F4 , G2 , G4 \} $ constitute a closed surface of genus zero in $SU(2) \sim S^3$.

As sketched in Figure~9, the set of the images of $\{ F2 , F4 , G2 , G4 \} $ can be globally parametrised by new variables $-1 \leq X \leq 1$ and $-1 \leq Y \leq 1$ according to the relations 
\bea 
\left [ G2 \right ] \; :  \qquad U_0 &=& e^{i \pidue (X+1) \sigma^2} e^{i \pidue Y \sigma^1} = e^{i \pidue X \sigma^2} e^{- i \pidue Y \sigma^1} \, i \sigma^2 = \widetilde U_0 \, i \sigma^2 \; , \nonumber \\ 
\left [ F4 \right ] \; :  \qquad U_0 &=& e^{- i \pidue Y \sigma^1} e^{i \pidue (1+ X) \sigma^2} = e^{-i \pidue Y \sigma^1} e^{i \pidue X \sigma^2} \, i \sigma^2 = \widetilde U_0 \, i \sigma^2\; , \nonumber \\ 
\left [ F2 \right ] \; :  \qquad U_0 &=& e^{- i \pidue Y \sigma^1} e^{i \pidue (1+ X) \sigma^2} = e^{-i \pidue Y \sigma^1} e^{i \pidue X \sigma^2} \, i \sigma^2 = \widetilde U_0 \, i \sigma^2\; , \nonumber \\ 
\left [ G4 \right ] \; :  \qquad U_0 &=& e^{i \pidue (X+1) \sigma^2} e^{i \pidue Y \sigma^1} = e^{i \pidue X \sigma^2} e^{- i \pidue Y \sigma^1} \, i \sigma^2 = \widetilde U_0 \, i \sigma^2 \; .  
\label{6.7}
\eea

\vskip 0.4 truecm

\centerline {
\begin{tikzpicture} [scale= 0.9] 
%
\draw [ -> ,  very thick  ] (2,0) -- (2,5);
\draw [ very thick ] (0,0) -- (0,4);
\draw [  very thick ] (4,0) -- (4,4);
\draw [ very thick ] (0,0) -- (4,0); 
\draw [ very thick ] (0,4) -- (4,4);
\draw [ -> ,  very thick  ] (0,2) -- (5,2);
\node at (1.6, 5) {$Y$}; 
\node at (5.4, 2) {$X$}; 
\node at (1, 3) {$G2$};
\node at (3, 3) {$F4$};
\node at (1, 1) {$F2$};
\node at (3, 1) {$G4$};
\node at (0, 0) {$\bullet$};
\node at (0, 2) {$\bullet$};
\node at (0, 4) {$\bullet$};
\node at (2, 0) {$\bullet$};
\node at (2, 2) {$\bullet$};
\node at (2, 4) {$\bullet$};
\node at (4, 0) {$\bullet$};
\node at (4, 2) {$\bullet$};
\node at (4, 4) {$\bullet$};
\node at (-0.4, 2.2) {$1$};
\node at (4.4, 2.2) {$-1$};
\node at (-0.4, -0.4) {$g_1$};
\node at (2, -0.4) {$g_1 g_2$};
\node at (4.4, -0.4) {$g_1$};
\node at (-0.4, 4.3) {$g_1$};
\node at (2.5, 4.3) {$g_2 g_1$};
\node at (4.4, 4.3) {$g_1$};
\node at (1, 4) {$/$};
\node at (3, 4) {$/$};
\node at (1, 0) {$/\! /$};
\node at (3, 0) {$/\!/$};
\node at (0, 3) {$\sim$};
\node at (0, 1) {$\sim$};
\node at (4, 3) {$\approx$};
\node at (4, 1) {$\approx$};
\end{tikzpicture}
}

\vskip 0.2 truecm

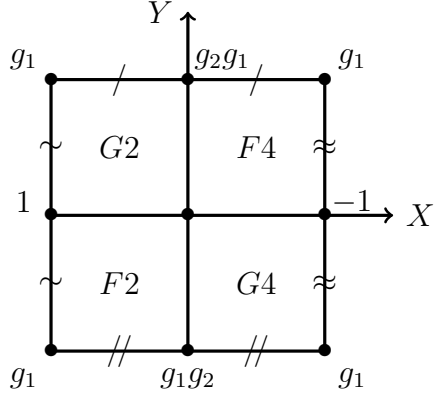
\centerline {{Figure~9.} {Images of the regions $\{ F2 , F4 , G2 , G4 \} $  parametrised in  equation (\ref{6.7}).}}

\vskip 0.4 truecm

\noindent The  images of $\{ F2 , F4 , G2 , G4 \} $  are glued as shown in Figure~9; the edges which are labelled by the same symbol must be identified.  Therefore, the closed surface which is  specified by $ \Phi_R^{-1} \Phi_{f*L} : \der H_R \rightarrow SU(2)$ is topologically equivalent to the tetrahedron shown in Figure~10.  Relations (\ref{6.7}) show that $U_0(X,Y)$  can globally be written as $U_0(X,Y)= \widetilde U_0(X,Y) \, i \sigma^2$, therefore if $\widetilde U$ denotes the extension of $\widetilde U_0$ in $H_R$, one has  
\be
\Gamma [U] = \Gamma [\widetilde U] \; . 
\label{6.8}
\ee
In order to determine the value of $\Gamma [\widetilde U]$ one can use symmetry arguments.

 \vskip 0.5 truecm 
  
\centerline {
\begin{tikzpicture} [scale= 0.9] 
%
\draw [  very thick  ] (0,0) -- (5,1);
\draw [ very thick ] (4,4) -- (1,5);
\draw [  very thick ] (0,0) -- (1,5);
\draw [ very thick ] (0,0) -- (4,4); 
\draw [ very thick ] (4,4) -- (5,1);
\draw [ very thick , dashed ] (1,5) -- (2.9,3.1);
\draw [ very thick , dashed ] (5,1) -- (3.1,2.9);
\node at (0, 0) {$\bullet$};
\node at (5, 1) {$\bullet$};
\node at (4, 4) {$\bullet$};
\node at (1, 5) {$\bullet$};
\node at (2.5, 0.5) {$\bullet$};
\node at (2.5, 4.5) {$\bullet$};
\node at (5.54, 1) {$-1$};
\node at (-0.4, -0.16) {$1$};
\node at (0.4, 5.4) {$g_2 g_1$};
\node at (2.6, 0.01) {$g_2$};
\node at (2.5, 4.9) {$g_1$};
\node at (4.6, 4.3) {$g_1g_2$};
\end{tikzpicture}
}

\vskip 0.2 truecm

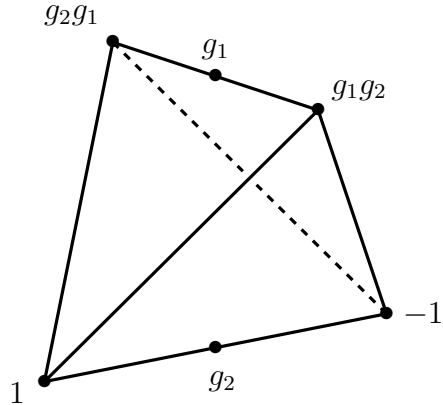
\centerline {{Figure~10.} {Closed surface specified by  $ \Phi_R^{-1} \Phi_{f*L} : \der H_R \rightarrow SU(2)$.}}

\vskip 0.4 truecm
 
\noindent The manifold $SU(2) \sim S^3$ can be represented as the union of two equivalent (with the same volume) balls  in ${\mathbb R}^3$ of radius $\pi/2$ with identified boundaries, $SU(2) \sim {\cal B}_1 \cup {\cal B}_2$. Indeed each element of $SU(2)$ can be written as 
$$
e^{i \ner \theta \ner \sigma} = \cos (|  \ner \theta | ) + i \widehat {\ner \theta} \ner \sigma \, \sin (| \ner \theta |) \; , 
$$
where $|\ner \theta | = [ \ner \theta \ner \theta ]^{1/2} $ and $ \widehat {\ner \theta}= (\ner \theta / |\ner \theta | )$.  The ball ${\cal B}_1$ contains the elements with $0 \leq |\ner \theta | \leq \pi / 2$, and ${\cal B}_2$ contains the elements with  $(\pi / 2) \leq | \ner \theta | \leq \pi $. 

The application $\widetilde U_0 : \der H_R \rightarrow SU(2)$ maps the boundaries of the rectangles $\{ F2 , F4 \} $ and $ \{ G2 , G4 \} $ into the eight   edges in ${\cal B}_1$ shown in Figure~11. Equation (\ref{6.7}) and the  picture of Figure~11 demonstrate that the surface $\widetilde U_0 : \der H_R \rightarrow SU(2)$ is symmetric under  rotations of $\pi /2 $ around the $\sigma^3$ axis and bounds a region $\cal R$ of $SU(2)$ which is contained in half of the ball ${\cal B}_1$. According to the reasoning of Section~\ref{RA}, the volume of this region $\cal R$ must take the value $n / 8$, where $n$ is an integer.  This integer $n$ is less than 4 because  $\cal R$ is contained inside ${\cal B}_1$ and satisfies $n \leq 2$ because $\cal R$ is contained inside half of  ${\cal B}_1$. Finally, the value $n=2 $ is excluded because a direct inspection shows that $\cal R$ does not cover the upper half-part of ${\cal B}_1$ completely. 
 Therefore one finally obtains 
\be
\Gamma [U] = \Gamma [\widetilde U] = {1\over 8} \; . 
\label{6.9}
\ee
In Section~8 it will be shown that equation (\ref{6.9}) is also in agreement with  a  direct computation of $\Gamma [U]$ that we have performed by means  of the canonical expression (\ref{4.18}). Finally, the validity of the  result (\ref{6.9}) has also been verified by means of a numerical evaluation of the integral (\ref{4.18}). To sum up, in the case of the manifold $\Sigma_3$ with the specified representation (\ref{6.2}) of its fundamental group, the value of the classical Chern-Simons invariant is given by 
\be
cs [A] = {1 \over 8} \quad \hbox{\rm mod ~}   {\mathbb Z} \; .
\label{6.10}
\ee

\vskip 0.4 truecm 

\centerline {
\begin{tikzpicture} [scale= 0.9] 
%
\draw [   ] (0.5,2) -- (2.85,2);
\draw [ ->  ] (3.2,2) -- (9.5,2);
\draw [ -> ] (5,2) -- (5,6);
\draw [ <-  ] (2,0.5) -- (8,3.5);
\draw [  very thick ]   (8,2) arc (0:180:3);
\draw [  very thick ]  (3,1) to [out=90, in=200] (5,5);
\draw [  very thick , dashed ]  (5,5) to [out=-8, in=97] (7,3);
\draw [very thick ] (2,2) -- (2.85,2); 
\draw [very thick ] (3.2,2) -- (8,2);
\draw [very thick ] (3,1) -- (7,3);
\node at (5, 2) {$\bullet$};
\node at (5, 5) {$\bullet$};
\node at (8, 2) {$\bullet$};
\node at (2, 2) {$\bullet$};
\node at (3, 1) {$\bullet$};
\node at (7, 3) {$\bullet$};
\node at (4.7, 2.3) {$1$};
\node at (1.4, 2.3) {$-i\sigma^2$};
\node at (8.5, 2.3) {$i\sigma^2$};
\node at (3, 0.5) {$i \sigma^1$};
\node at (6.2, 3.2) {$-i \sigma^1$};
\node at (4.6, 5.4) {$i \sigma^3$};
\end{tikzpicture}
}

\vskip 0.4 truecm

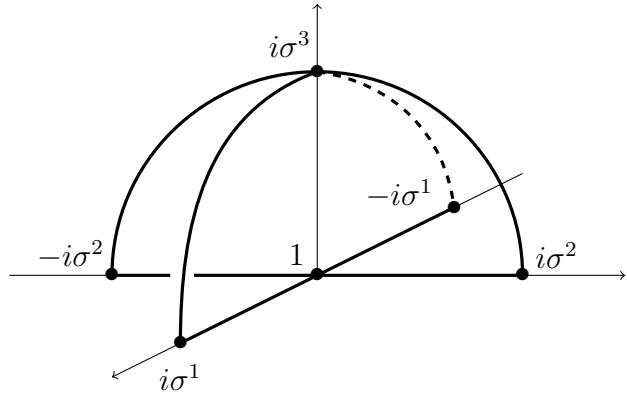
\centerline {{Figure~11.} {$\widetilde U_0$ images in ${\cal B}_1$ of the boundaries of the regions  $\{ F2 , F4 , G2 , G4 \} $}.}

\vskip 0.3 truecm

\section{Poincar\'e sphere}

The Poincar\'e sphere $\cal P$ admits a genus 2 Heegaard splitting presentation.   The corresponding   Heegaard diagram  \cite{R}  is  shown  in Figure~12.      One of the two characteristic curves, $C^\prime_1 = f(C_1)$,  is described by the continuous line, whereas the second curve $C^\prime_2 = f(C_2)$ is represented by the  dashed path; $x_b$ designates the base point for the fundamental group. 

Let the generators $\{ \gamma_1 , \gamma_2 \} $  of $\pi_1(H_R)$ be  associated with $+F$ and $+G$ respectively and oriented in the outward direction,  According to  the Heegaard diagram of  Figure~12,   the homotopy class of $C^\prime_1$ is given by $\gamma_1^{-4} \gamma_2 \gamma_1  \gamma_2 $, whereas the class of $C_2^\prime$ is equal to $\gamma_2^{-2} \gamma_1 \gamma_2  \gamma_1 $. Therefore the fundamental group of $\cal P$ admits  the presentation 
 \be
 \pi_1({\cal P}) = \langle \, \gamma_1, \gamma_2   \, | \, \gamma_1^5 =  \gamma_2^3 = (\gamma_1 \gamma_2)^2 \, \rangle \; ,   
 \label{7.1}
 \ee
 which corresponds to the binary icosahedral (or dodecahedral) group of order $120$.  Since the abelianization of $\pi_1({\cal P})$ is trivial, $\cal P$ is a homology sphere. 
A nontrivial representation  $\rho : \pi_1({\cal P}) \rightarrow SU(2)$ is given \cite{LES,GUM}  by 
\bea
\rho(\gamma_1) &=& g_1 = e^{i b_1} =\exp \left [ i {\pi \over 5} \sigma  \right ] \; , \nonumber \\
\rho(\gamma_2) &=& g_2 = e^{i b_2} = \exp \left [ i {\pi \over 3} \widetilde \sigma  \right ] \; ,  
\label{7.2}
\eea
where 
\bea
\sigma &=& \begin{pmatrix} 1 & 0 \\ 0 &-1 \end{pmatrix} \; , \nonumber \\
\widetilde \sigma &=& r \begin{pmatrix} 1 & 0 \\ 0 &-1 \end{pmatrix} + \sqrt {1-r^2} \begin{pmatrix} 0 & 1 \\ 1 &0 \end{pmatrix} \; , \nonumber \\
r &=& {\cos (\pi /3) \, \cos (\pi /5) \over \sin (\pi /3) \, \sin (\pi /5)} \; . 
\label{7.3}
\eea

\vskip 0.5 truecm

\centerline {
\begin{tikzpicture} [scale=0.64] 
%
\draw [very thick 
] (7,10) circle (1.7); 
\draw [very thick 
] (7,5) circle (1.7); 
\draw [very thick 
] (1,1) circle (1.7); 
\draw [very thick 
] (13,1) circle (1.7); 
\draw [  very thick , densely dashed 
] (7,8.3) to [out=-90 , in=90 ] (7,6.7);
\draw [ very thick , densely dashed
] (0.2,2.49) to [out=90 , in=200 ] (5.5,10.8);
\draw [  very thick
] (1,2.7) to [out=78 , in=215 ] (5.3,9.8);
\draw [  very thick , densely dashed 
] (1.68,2.54) to [out=60 , in=190 ] (5.31,5.1);
\draw [  very thick
] (2.25,2.16) to [out=24 , in=227 ] (5.57,4.11);
\draw [  very thick
] (2.62,1.5) to [out=10 , in=170 ] (11.4,1.5);
\draw [  very thick
] (2.7,0.8) to [out=2 , in=178 ] (11.3,0.8);
\draw [  very thick
] (2.45,0.15) to [out=-9 , in=188 ] (11.52,0.15);
\draw [  very thick , densely dashed
] (11.7,2.07) to [out=160 , in=-45 ] (8.45,4.08);
\draw [ very thick 
] (12.31,2.53) to [out=130 , in=-20 ] (8.7,5.2);
\draw [  very thick , densely dashed
] (13,2.7) to [out=100 , in=-40 ] (8.67,9.77);
\draw [  very thick
] (13.8,2.48) to [out=90 , in=-20 ] (8.48,10.81);
\node at (7,10) {$  - G $};
\node at (7,5) {$  + G$};
\node at (1,1) {$  + F $};
\node at (13,1) {$  - F $};
\node at (8.15,10.80) {\footnotesize $  1 $};
\node at (5.89,10.80) {\footnotesize $  5 $};
\node at (8.27,9.80) {\footnotesize $  4 $};
\node at (5.73,9.80) { \footnotesize $ 2 $};
\node at (7,8.7) {\footnotesize $  3 $};
\node at (7,6.3) {\footnotesize $  4 $};
\node at (8.3, 5.3) {\footnotesize $  1 $};
\node at (8.2,4.3) {\footnotesize $  5 $};
\node at (5.81,4.3) {\footnotesize $  2 $};
\node at (5.8,5.2) {\footnotesize $  3 $};
\node at (0.2,2.1) {\footnotesize $  6 $};
\node at (1,2.3) {\footnotesize $  5 $};
\node at (1.5,2.2) {\footnotesize $  7 $};
\node at (1.9,1.99) {\footnotesize $  1 $};
\node at (2.25,1.45) {\footnotesize $  2 $};
\node at (2.3,0.8) {\footnotesize $  3 $};
\node at (2.1,0.22) {\footnotesize $  4 $};
\node at (11.9,0.22) {\footnotesize $  3 $};
\node at (11.7,0.8) {\footnotesize $  2 $};
\node at (11.7,1.45) {\footnotesize $  1 $};
\node at (12,1.91) {\footnotesize $  7 $};
\node at (12.44,2.28) {\footnotesize $  5 $};
\node at (13,2.4) {\footnotesize $  6 $};
\node at (13.7,2.2) {\footnotesize $  4 $};
\node at (3.2,3.4) {$\bullet$};
\node at (3.7,3.7) {$x_b$};
\node at (12.4,9) {$C_1^\prime$}; 
\node at (-0.01,6) {$C_2^\prime$}; 
\draw [-> , very thick ] (-0.2,-0.2) -- ++(140:1mm); 
\draw [-> , very thick ] (14.2,-0.2) -- ++(50:1mm); 
\draw [-> , very thick ] (7,3.3) -- ++(180:1mm); 
\draw [-> , very thick ] (7,11.7) -- ++(180:1mm); 
\draw [-> , very thick ] (1,6.6) -- ++(68:1mm); 
\draw [-> , very thick ] (2.4,6.71) -- ++(62:1mm); 
\draw [-> , very thick ] (3.1,4.2) -- ++(38:1mm); 
\draw [-> , very thick ] (4,2.94) -- ++(211:1mm); 
\draw [-> , very thick ] (7,1.94) -- ++(180:1mm); 
\draw [-> , very thick ] (7,0.89) -- ++(180:1mm); 
\draw [-> , very thick ] (7,-0.24) -- ++(180:1mm); 
\draw [-> , very thick ] (10,2.84) -- ++(-36:1mm); 
\draw [-> , very thick ] (10.95,3.94) -- ++(-43:1mm); 
\draw [-> , very thick ] (7,7.5) -- ++(-90:1mm); 
\draw [-> , very thick ] (11.82,6.21) -- ++(-59:1mm);
\draw [-> , very thick ] (12.82,6.99) -- ++(118:1mm);
\end{tikzpicture}
}

\vskip 0.5 truecm
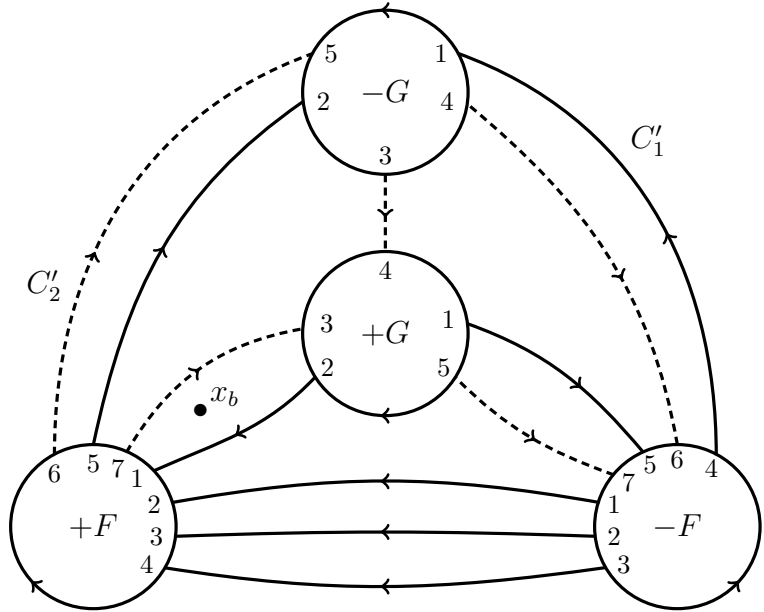
\centerline {{Figure~12.} {Heegaard diagram for the Poincar\'e sphere.}}

\vskip 0.7 truecm

Equation (\ref{7.2}) specifies the values of  $A^0_R$,  
  \be
 A^0_R  =  \left  \{ \begin{array}  {l@{ ~ } l}  
  b_1  \, \theta^\prime (t_1) dt_1 & \quad \hbox{ inside a neighbourhood of } +F   \; ;    \\   ~ & ~ \\  
b_2 \,  \theta^\prime (t_2) dt_2 &  \quad \hbox{ inside a neighbourhood of  }  +G \; ; \\ ~ & ~ \\  0 & \quad \hbox { otherwise } \; . 
\end{array} \right. 
\label{7.4}
\ee
The values of $A_L^0$ are  determined by equation (\ref{7.2}) and by the choice of the base point. Indeed, let    the  generators   $\{ \lambda_1 , \lambda_2 \}$   of $\pi_1(H_L)$ be associated with $C_1$ and $C_2$ respectively. Then, from the Heegaard diagram and  the position for the base point, one finds 
\bea
\rho(\lambda_1) &=& g_1 = e^{i b_1} = \exp \left [ i {\pi \over 5} \sigma  \right ] \; , \nonumber \\
\rho(\lambda_2) &=& g_2 = e^{i b_2} = \exp \left [ i {\pi \over 3} \widetilde \sigma  \right ] \; . 
\label{7.5}
\eea
Consequently, the image of $A^0_L$ under the gluing homeomorphism $f$ takes values 
  \be
 f*A^0_L  =  \left  \{ \begin{array}  {l@{ ~ } l}  
  b_1 \, \theta^\prime (u_1) du_1 & \quad \hbox{ inside a neighbourhood of } C_1^\prime   \; ;    \\   ~ & ~ \\  
b_2 \,  \theta^\prime (u_2) du_2 &  \quad \hbox{ inside a neighbourhood of  }  C_2^\prime \; ; \\ ~ & ~ \\  0 & \quad \hbox { otherwise } \; . 
\end{array} \right. 
\label{7.6}
\ee

\vskip 0.4 truecm

\centerline {
\begin{tikzpicture} [scale=0.68] 
%
\draw [very thick 
] (7,10) circle (1.7); 
\draw [very thick 
] (7,5) circle (1.7); 
\draw [very thick 
] (1,1) circle (1.7); 
\draw [very thick 
] (13,1) circle (1.7); 
\draw [  very thick , densely dashed 
] (7,8.3) to [out=-90 , in=90 ] (7,6.7);
\draw [ very thick , densely dashed
] (0.2,2.49) to [out=90 , in=200 ] (5.5,10.8);
\draw [  very thick
] (1,2.7) to [out=78 , in=215 ] (5.3,9.8);
\draw [  very thick , densely dashed 
] (1.68,2.54) to [out=60 , in=190 ] (5.31,5.1);
\draw [  very thick
] (2.25,2.16) to [out=24 , in=227 ] (5.57,4.11);
\draw [  very thick
] (2.62,1.5) to [out=10 , in=170 ] (11.4,1.5);
\draw [  very thick
] (2.7,0.8) to [out=2 , in=178 ] (11.3,0.8);
\draw [  very thick
] (2.45,0.15) to [out=-9 , in=188 ] (11.52,0.15);
\draw [  very thick , densely dashed
] (11.7,2.07) to [out=160 , in=-45 ] (8.45,4.08);
\draw [ very thick 
] (12.31,2.53) to [out=130 , in=-20 ] (8.7,5.2);
\draw [  very thick , densely dashed
] (13,2.7) to [out=100 , in=-40 ] (8.67,9.77);
\draw [  very thick
] (13.8,2.48) to [out=90 , in=-20 ] (8.48,10.81);
\node at (7,10) {$  - G $};
\node at (7,5) {$  + G$};
\node at (1,1) {$  + F $};
\node at (13,1) {$  - F $};
\node at (8.15,10.80) {\footnotesize $  1 $};
\node at (5.89,10.80) {\footnotesize $  5 $};
\node at (8.27,9.80) {\footnotesize $  4 $};
\node at (5.73,9.80) { \footnotesize $ 2 $};
\node at (7,8.7) {\footnotesize $  3 $};
\node at (7,6.3) {\footnotesize $  4 $};
\node at (8.3, 5.3) {\footnotesize $  1 $};
\node at (8.2,4.3) {\footnotesize $  5 $};
\node at (5.81,4.3) {\footnotesize $  2 $};
\node at (5.8,5.2) {\footnotesize $  3 $};
\node at (0.2,2.1) {\footnotesize $  6 $};
\node at (1,2.3) {\footnotesize $  5 $};
\node at (1.5,2.2) {\footnotesize $  7 $};
\node at (1.9,1.99) {\footnotesize $  1 $};
\node at (2.25,1.45) {\footnotesize $  2 $};
\node at (2.3,0.8) {\footnotesize $  3 $};
\node at (2.1,0.22) {\footnotesize $  4 $};
\node at (11.9,0.22) {\footnotesize $  3 $};
\node at (11.7,0.8) {\footnotesize $  2 $};
\node at (11.7,1.45) {\footnotesize $  1 $};
\node at (12,1.91) {\footnotesize $  7 $};
\node at (12.44,2.28) {\footnotesize $  5 $};
\node at (13,2.4) {\footnotesize $  6 $};
\node at (13.7,2.2) {\footnotesize $  4 $};
\node at (3.3,3.4) {$1$};
\node at (4,6) {$g_2$}; 
\node at (2.07,7.3) {$g_2 g_1$}; 
\node at (1.3,9) {$g_1^4$};
\node at (9,7) {$g^2_2$};
\node at (11.9,7.3) {$-1$}; 
\node at (10.2,3.5) {$g_1 g_2$};
\node at (5.5,2.6) {$g_1$}; 
\node at (5.8,1.4) {$g^2_1$}; 
\node at (6.3,0.4) {$g^3_1$}; 
\draw [-> , very thick ] (-0.2,-0.2) -- ++(140:1mm); 
\draw [-> , very thick ] (14.2,-0.2) -- ++(50:1mm); 
\draw [-> , very thick ] (7,3.3) -- ++(180:1mm); 
\draw [-> , very thick ] (7,11.7) -- ++(180:1mm); 
\draw [-> , very thick ] (1,6.6) -- ++(68:1mm); 
\draw [-> , very thick ] (2.4,6.71) -- ++(62:1mm); 
\draw [-> , very thick ] (3.1,4.2) -- ++(38:1mm); 
\draw [-> , very thick ] (4,2.94) -- ++(211:1mm); 
\draw [-> , very thick ] (7,1.94) -- ++(180:1mm); 
\draw [-> , very thick ] (7,0.89) -- ++(180:1mm); 
\draw [-> , very thick ] (7,-0.24) -- ++(180:1mm); 
\draw [-> , very thick ] (10,2.84) -- ++(-36:1mm); 
\draw [-> , very thick ] (10.95,3.94) -- ++(-43:1mm); 
\draw [-> , very thick ] (7,7.5) -- ++(-90:1mm); 
\draw [-> , very thick ] (11.82,6.21) -- ++(-59:1mm);
\draw [-> , very thick ] (12.82,6.99) -- ++(118:1mm);
\end{tikzpicture}
}

\vskip 0.5 truecm
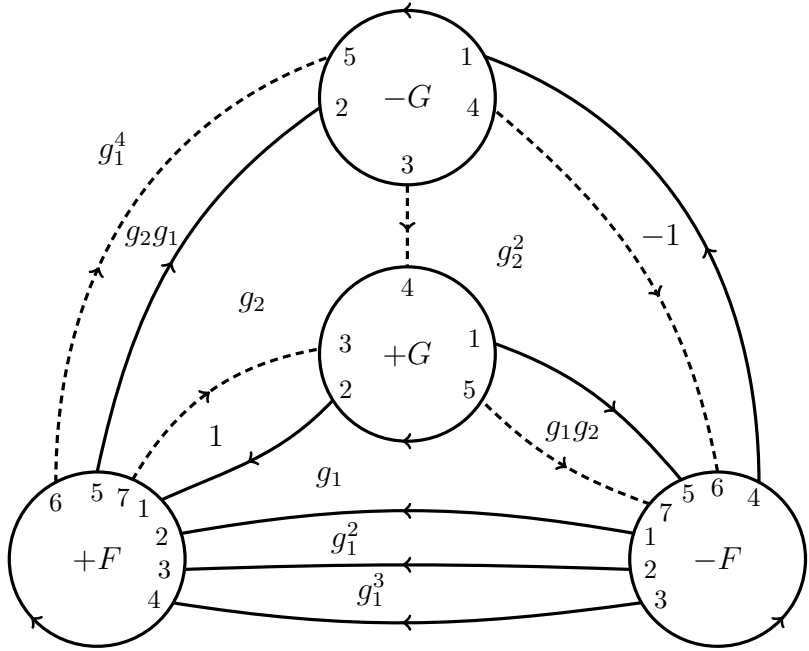
\centerline {{Figure~13.} {Values of $U_0$ in the region where  $f*A^0_L$ and $A^0_R$ vanish.}}

\vskip 0.7 truecm

One can now determine the map $U_0 = \Phi_R^{-1} \Phi_{f*L} : \der H_R \rightarrow SU(2)$. In the region of the surface $\der H_R$ where both $f*A^0_L$ and $A^0_R$ are vanishing, the values of $U_0$ are shown in Figure~13. By using the method illustrated in the previous examples, one can compute the classical Chern-Simons invariant. The intersection component is given by 
\bea
{\cal X} [A] &=& \frac{1}{8 \pi^2 }\Bigl \{  -4 \, {\rm Tr} \left ( b_1 b_1 \right ) -2 \, {\rm Tr} \left ( b_2 b_2 \right ) 
+ 4 \, {\rm Tr} \left ( b_1 b_2 \right ) \nonumber \\
&& {\hskip 2 cm} + {\rm Tr} \left ( b_1 g_2b_1 g_2^{-1}\right ) + {\rm Tr} \left ( b_2 g_1b_2 g_1^{-1}\right ) \Bigr \} \nonumber \\
&=& - {2\over 15}  + {1\over 2} \left [ {1\over 5}{\cos (\pi / 3)\over \sin(\pi / 5)} + {1\over 3} {\cos(\pi /5) \over \sin (\pi /3) }     \right ]^2
\; . 
\label{7.7}
\eea 
The image of the map $\Phi_R^{-1} \Phi_{f*L} : \der H_R \rightarrow SU(2)$ is a genus 0 surface in the group $SU(2)$. We skip the details, which anyway can be obtained from the Heegaard diagram and equations (\ref{7.2})-(\ref{7.6}).   Numerical computations of the integral (\ref{4.18}) give the following value of the Wess-Zumino volume (with $10^{-10}$ precision)
\be
\Gamma [A] = 0.0090687883 \cdots  \; . 
\label{7.8}
\ee
 Therefore, the value of the classical Chern-Simons invariant associated with the representation (\ref{7.2}) of $\pi_1 ({\cal P})$ turns out to be 
\be
cs[A] = - 0.0083333333 \cdots = - {1\over 120} \quad \hbox{\rm mod ~}   {\mathbb Z} \; , 
\label{7.9}  
\ee
where the last identity is a consequence of the fact that $\left | \pi_1 ({\cal P}) \right | = 120$. 
The result (\ref{7.9}) has also been obtained by means of a complete computation of the integral (\ref{4.18}); this issue is elaborated in Section~8.  

\section{Computations of the Wess-Zumino volume}

The computation of $\Gamma [U]$ by means of the canonical expression (\ref{4.18}) presents  general features that are consequences of  our construction of the flat connection $A$ 
by means of a Heegaard splitting presentation of $M$. This allows the derivation of  universal  formulae of the classical Chern-Simons invariant for quite wide classes of manifolds. 
 We present here one example; details will be produced in a forthcoming article. 
 
Let us consider the set of Seifert spaces $ \Sigma(m,n,-2) $  of genus zero with three singular fibers which are characterised by the integer surgery coefficients $(m,1)$, $(n,1)$ and $(2,-1)$.  The manifolds $\Sigma(m,n,-2)$  admit  \cite{S,LES} a genus two Heegaard splitting $M = H_L \cup_f H_R$ and  their fundamental group can be presented as  
\be
 \pi_1(M) = \langle \, \gamma_1, \gamma_2   \, | \, \gamma_1^m =  \gamma_2^n = (\gamma_1 \gamma_2)^2 \, \rangle \; ,   
 \label{8.1}
\ee
for nontrivial positive integers $m$ and $n$. The manifold $\Sigma_3$ discussed in Section~6 and the Poincar\'e manifold $\cal P$ considered in Section~7 are examples belonging to this class of manifolds. Let us introduce the representation of $\pi_1(M)$ in the group $SU(2)$ given by 
\bea
\gamma_1 &\rightarrow& g_1 = \exp \left [ i \theta_1 \sigma \, \right ]  \quad , \nonumber \\ 
\gamma_2 &\rightarrow& g_2 = \exp \left [ i \theta_2 \widetilde \sigma \, \right ]  \quad ,  
\label{8.2}
\eea
where $\sigma$ and $\widetilde \sigma$ are combinations of the sigma matrices satisfying $\sigma^2 = 1 = \widetilde \sigma^2 $, and  
\be
g_1^m = g_2^n = \left ( g_1 g_2 \right )^2 = -1   \; . 
\label{8.3}
\ee
In this case, the value of the surface integral (\ref{4.5}) is given by 
\be
{\cal X}[A] = - {1\over 4} \left \{ m \left [ {\theta_1\over \pi} \right ]^2 + n \left [ {\theta_2\over \pi} \right ]^2 -2 \left [ \left ( \theta_2 \over \pi \right ){\cos \theta_1 \over \sin \theta_2} +    \left ( \theta_1 \over \pi \right ){\cos \theta_2 \over \sin \theta_1}\right ]^2 \right \} \; . 
\label{8.4}
\ee
As it has been shown in the previous examples, the image of the map $\Phi_R^{-1} \Phi_{f*L} : \der H_R \rightarrow SU(2)$ is a genus 0 surface in the group $SU(2)$. 
The corresponding Wess-Zumino volume turns out to be 
\be
\Gamma [U] =  {1\over 4} \left \{ {1\over 2}  -2 \left [ \left ( \theta_2 \over \pi \right ){\cos \theta_1 \over \sin \theta_2} +    \left ( \theta_1 \over \pi \right ){\cos \theta_2 \over \sin \theta_1}\right ]^2 \right \} \; . 
\label{8.5}
\ee
 So that the value of the classical Chern-Simons invariant for the manifolds $\Sigma (m,n,-2)$ reads 
\be
cs[A] = - {1\over 4} \left \{  m \left [ {\theta_1\over \pi} \right ]^2 + n \left [ {\theta_2\over \pi} \right ]^2  - {1\over 2 } \right \} \quad \hbox{mod ~}   {\mathbb Z}   \; . 
\label{8.6}
\ee
When $m=n=2$,   expression (\ref{8.6}) gives  the value of the classical Chern-Simons invariant appearing in equation (\ref{6.9}); and for $m=5, n=3$, expression (\ref{8.6}) coincides with equation (\ref{7.9}). Equation (\ref{8.6}) is valid for generic values of $m$ and $n$; for those particular values of $m$ and $n$   such that $\Sigma (m,n,-2)$ is a Seifert homology sphere,  our equation (\ref{8.6}) is in agreement with the results of Fintushel and Stern \cite{FI} and Kirk and Klassen \cite{KK} for  Seifert spheres. 

\section{Conclusions}

Given a $SU(N)$ representation $\rho $ of the fundamental group of a 3-manifold $M$,  we have shown how to define a corresponding flat connection $A$ on $M$ such that the holonomy of $A$ coincides with $\rho$. Our construction is based on a Heegaard splitting presentation of $M$, so that the relationship  between $A$ and the topology of $M$ is  displayed.  The relative classical Chern-Simons invariant $cs[A]$ is naturally decomposed into the sum of two contributions: a sort of coloured intersection form, which is specified by the Heegaard diagram, and a Wess-Zumino volume of a region of $SU(N)$ which is determined by the non commutative structure of the $\rho $ representation   of $\pi_1(M)$. 
A canonical expression for the Wess-Zumino  volume, as function of the boundary data exclusively,    has been produced.  A few illustrative examples of  flat connections and of classical Chern-Simons invariant computations  have been presented. 

\vskip 0.8 truecm 

\noindent {\large \bf Acknowledgments.} We wish to thank R.~Benedetti and C.~Lescop for discussions. 

\vskip 1.3 truecm 

\bibliographystyle{amsalpha}

\end{document}